%% file: main.tex
\documentclass[aip,cha,reprint,floatfix]{revtex4-1}

\usepackage{graphicx}
\usepackage{amsmath,mathrsfs}
\usepackage{bm}
\usepackage{siunitx}
\usepackage[hidelinks]{hyperref}
\usepackage{subcaption}

\newcommand{\Temp}{\mathscr{T}}

\begin{document}

\title{Planar Lagrangian transport and scalar-gradient organization in a turbulent reacting shear layer}

\author{Sriram P. Kalathoor}
\affiliation{Daniel Guggenheim School of Aerospace Engineering, Georgia Institute of Technology, Atlanta GA 30332}

\author{Joseph C. Oefelein}
\affiliation{Daniel Guggenheim School of Aerospace Engineering, Georgia Institute of Technology, Atlanta GA 30332}

\date{\today}

\begin{abstract}
We analyze planar Lagrangian transport and scalar-gradient organization in a supersonic, reacting hydrogen--air temporal mixing layer using time-resolved mid-plane data from a three-dimensional direct numerical simulation. The analysis combines forward/backward finite-time Lyapunov exponent (FTLE) fields, operational FTLE-ridge skeletons, Cauchy--Green deformation measures, shear-LCS metrics, and planar hyperbolic geodesic-LCS extraction to examine how finite-time stretching structures the reacting shear layer. The time-resolved FTLE ridges identify repelling and attracting finite-time transport skeletons in the constrained two-dimensional slice, from which ridge geometry, intersection occupancy, persistence, and scalar-conditioned transport are quantified. Hyperbolic geodesic LCS are extracted from Cauchy--Green tensors reconstructed from planar flow maps as strainlines seeded at high-$\lambda_{\max}$ normal maxima, providing a variational counterpart to the operational FTLE-ridge skeleton. We then relate the transport skeleton to temperature, mixture fraction, and a reaction intermediate. The results show localized forward/backward ridge overlap, strong scalar-gradient enrichment, finite-time geodesic LCS that occupy the same high-strain transport skeleton, residual direction-dependent separation from a time- and cross-stream-stratified null model, and scalar-response lags that remain compact relative to decorrelation and FTLE-integration scales. Together, these results provide a transport-oriented characterization of coherent structures and their role in mid-plane mixing within a compressible reacting shear flow.
\end{abstract}

\maketitle

\begin{quotation}
Coherent structures in turbulent reacting flows control where material is stretched, folded, and mixed, but they are difficult to identify from instantaneous snapshots alone. Finite-time Lagrangian transport analysis of a planar slice through a supersonic reacting mixing layer locates attracting and repelling transport skeletons and relates them to thermal, mixture-fraction, and intermediate-species gradients. The results show that the transport skeleton is directionally asymmetric, concentrated into recurrent hotspots, and strongly enriched in scalar gradients even after accounting for shear-layer position. The study provides a nonlinear-dynamics view of how finite-time transport organizes scalar structure in a compressible reacting flow.
\end{quotation}

\section{Introduction}
Lagrangian coherent structures (LCS) and related finite-time Lagrangian methods provide a framework for identifying transport-organizing structures in unsteady flows \citep{Haller2015}. By tracking finite-time trajectory deformation, Lagrangian analysis exposes barriers and pathways for transport and mixing \citep{Shadden2005}. Hyperbolic transport skeletons are often approximated from forward/backward FTLE ridges, and variational geodesic formulations provide a more restrictive invariant description of coherent material barriers \citep{HallerBeronVera2012}. Experimental and data-driven analyses further show that these structures partition dynamically distinct flow regions in practice \citep{Kelley2013,Mathur2007}.

For finite-time trajectory maps, the FTLE is built from the largest eigenvalue of the right Cauchy--Green tensor and quantifies local exponential separation over a finite horizon \citep{Shadden2005,Haller2015}. Mathematically, for initial position $\mathbf{x}_0$ and window $T$,
\begin{equation}
\sigma_T(\mathbf{x}_0) = \frac{1}{2|T|} \ln \lambda_{\max}(\mathbf{x}_0, T),\label{eqn:ftle}
\end{equation}
where $\lambda_{\max}$ is the maximum eigenvalue of $\mathbf{C}=\mathbf{F}^{T}\mathbf{F}$, with $\mathbf{F}$ the finite-time deformation gradient, and $\mathbf{F}^T$ its transpose \citep{Haller2001}. In this use, FTLE ridges mark regions of strongest finite-time stretching or compression and approximate repelling/attracting transport skeletons \citep{Shadden2005,Haller2015,AllshousePeacock2015}. We use these FTLE ridges as operational, time-resolved transport skeletons and separately extract planar hyperbolic geodesic LCS from reconstructed Cauchy--Green eigenstructure over a sampled set of start times.

Lagrangian transport analyses are also useful in reacting-flow settings, where deformation and finite-time stirring strongly influence scalar segregation and reaction-zone organization. Prior combustion-instability studies have shown that Lagrangian structure measures can reveal coherent pathways that are difficult to infer from instantaneous Eulerian fields alone \citep{Sampath2016}. More broadly, coherent structures in turbulent scalar fields concentrate scalar-mixing support and large-gradient regions into nonuniform patterns \citep{Frohlich2008,Eisma2021,VanderwelTavoularis2016}. The same physics framework can probe high-speed reacting shear flows, where transport, mixing, and reaction remain tightly coupled.

Supersonic mixing layers are canonical compressible shear flows with strong coupling between vortical motion, compressibility, and scalar transport \citep{PapamoschouRoshko1988}. These characteristics make them a natural target for finite-time Lagrangian analysis, since forward/backward stretching fields provide a direct transport-oriented view complementary to standard Eulerian descriptions \citep{Shadden2005,Haller2015}.

Over the past decade, LCS methodology has expanded substantially through comparative evaluations of detection approaches \citep{AllshousePeacock2015,Hadjighasem2017}, objective Eulerian coherent-structure measures and vortex criteria \citep{SerraHaller2016,Hadjighasem2016}, transport-barrier formulations for diffusive/stochastic settings \citep{HallerKarraschKogelbauer2018,KarraschKeller2020}, compressible-flow FTLE analyses \citep{Gonzalez2016}, and uncertainty-aware applications in geophysical forecasting \citep{Serra2020SAR,Matuszak2025}. Recent work also extends coherent-structure analyses to increasingly complex flow-physics settings \citep{SiFang2024}. For high-speed reacting flows, where compressibility, finite-rate chemistry, and strong shear coexist, these developments motivate joint assessment of transport-skeleton geometry, scalar transport, sensitivity to extraction choices, null-model separation, and relevant temporal scales.

In this paper, we analyze planar Lagrangian transport and scalar-gradient organization in a supersonic, reacting hydrogen--air temporal mixing layer (TML), with sensitivity and temporal-consistency analyses appropriate to a finite-time, planar extraction. The study is organized around three physics questions. First, how asymmetric and localized are the attracting and repelling transport skeletons in the mid-plane, and do geodesic strainlines extracted from reconstructed Cauchy--Green tensors occupy the same high-strain scaffold? Second, how strongly do these structures condition thermal and species-gradient support once shear-layer position is controlled through a time- and cross-stream-stratified null model? Third, do the observed flow-scalar relationships persist under FTLE-ridge, integration-window, and geodesic-extraction sensitivity variations, and are their lead/lag scales compact relative to decorrelation and FTLE-integration times? All conclusions are drawn at the planar level, so care is required when connecting the mid-plane transport skeleton to fully three-dimensional material surfaces and to the broader reacting-flow dynamics.

\section{Methodology}

\subsection{Flow configuration and planar data}
The source data come from direct numerical simulation (DNS) of a three-dimensional supersonic turbulent reacting hydrogen--air temporal mixing layer following the configuration of \citet{OBrien2014}. The full compressible governing equations for mass, momentum, total energy, species, and thermodynamic state are integrated in time; numerical details follow \citet{Oefelein2006-PAS}. The streamwise and spanwise directions are periodic, and the present analysis uses the mid-$z$ plane as a time-resolved section through the three-dimensional field.

The upper stream is hot air/oxidizer with $M_1=2.25$, $U_1\approx1.75~\mathrm{km\,s^{-1}}$, $T_1=1500~\mathrm{K}$, mole fractions $(X_{\mathrm{O}_2},X_{\mathrm{N}_2})=(0.21,0.79)$, and mixture fraction $Z=0$. The lower stream is cold nitrogen-diluted hydrogen with $M_2=2.73$, $U_2\approx-1.22~\mathrm{km\,s^{-1}}$, $T_2=500~\mathrm{K}$, mole fractions $(X_{\mathrm{H}_2},X_{\mathrm{N}_2})=(0.06,0.94)$, and $Z=1$. Both streams are initialized at $p=202.65~\mathrm{kPa}$, and the reference length is the initial vorticity thickness $\delta_{\omega,0}$.

The present analysis uses time-resolved mid-plane fields from the reacting TML, with velocity components $u(x,y,t)$ and $v(x,y,t)$ used for planar trajectory advection and scalar fields ($\Temp$, $Z$, and HO$_2$ mass fraction) used for coupling measures. Spatial coordinates in the figures are nondimensionalized as $x/L_x$ and $y/L_y$ using the streamwise and layer-normal extents of the sampled plane. The underlying flow evolves in $(x,y,z,t)$, and the present work constrains the Lagrangian calculation to an in-plane surrogate system on the mid-$z$ slice. The out-of-plane velocity component $w$ is excluded from trajectory advection and retained as a measure of how strongly the analyzed slice samples three-dimensional motion. Table~\ref{tab:out_of_plane_velocity} reports the resulting RMS ratios. The results constitute a two-dimensional transport analysis of a three-dimensional reacting flow, with care taken when connecting mid-plane structures to fully three-dimensional material surfaces.

\input{tables/out_of_plane_velocity_summary.tex}

\subsection{FTLE computation}
Forward and backward FTLE fields are computed from finite-time flow maps by integrating particle trajectories over an integration window. The window labels used below denote relative finite-time horizons; the FTLE normalization in Eq.~\ref{eqn:ftle} uses the corresponding physical time interval from the stored data. The trajectory integration uses bilinear spatial interpolation, periodic wrapping in $x$, clipping at the $y$ boundaries, and four explicit substeps per stored velocity interval. Velocity values within those substeps are linearly interpolated between adjacent stored slices. For each analysis start time, the deformation gradient $\mathbf{F}$ is estimated from the flow map, and the right Cauchy--Green tensor $\mathbf{C}=\mathbf{F}^{T}\mathbf{F}$ yields $\lambda_{\max}$ for Eq.~\ref{eqn:ftle}. The main forward/backward FTLE evaluations use the longer reference window, and sensitivity variants at other window lengths are used to assess window dependence.

\subsection{Ridge extraction and statistics}
Operational FTLE ridges are extracted by thresholding each forward/backward FTLE sequence at a sequence-level cutoff. The cutoff is the median of the per-frame percentile values. Unless otherwise stated, the percentile is 95\%, with 93\% used as a threshold-sensitivity case. The resulting high-FTLE masks are skeletonized and small connected components are removed. Ridge length is then estimated from skeleton arclength using the local grid spacing, and ridge strength is the FTLE value sampled on the skeleton support. This produces binary ridge masks for each time, as well as aggregate length distributions, strength distributions, spatial occupancy maps, and streamwise column-integrated occupancy profiles. Forward and backward ridge populations are further compared through combined PDFs and normalized profiles.

\subsection{Planar hyperbolic geodesic-LCS extraction}
Planar hyperbolic geodesic LCS are extracted from the two-dimensional Cauchy--Green tensor field \citep{FarazmandHaller2012,Haller2015}. The representative visualization uses the mid-window tensor field. The time-resolved geodesic statistics reconstruct Cauchy--Green tensors directly from the full planar velocity sequence at 55 start times per direction spanning the synchronized interval. The denser operational FTLE-ridge statistics use every synchronized ridge mask. Let $\lambda_1\le\lambda_2$ and $\boldsymbol{\xi}_1,\boldsymbol{\xi}_2$ denote the eigenvalues and unit eigenvectors of $\mathbf{C}$. Repelling hyperbolic geodesic LCS in the forward-time field are extracted as strainlines tangent to the weakest-stretching eigenvector,
\begin{equation}
\frac{d\mathbf{r}}{ds}=\boldsymbol{\xi}_1(\mathbf{r}),
\end{equation}
seeded where the strongest stretching eigenvalue is a local maximum in the normal eigenvector direction,
\begin{equation}
\partial_{\boldsymbol{\xi}_2}\lambda_2=0,\qquad
\partial_{\boldsymbol{\xi}_2}^2\lambda_2<0,\qquad
\lambda_2>1.
\end{equation}
Backward-time Cauchy--Green fields are processed with the same criterion to obtain the attracting family. Numerically, seeds are selected from the upper tail of $\lambda_2$ normal maxima, strainlines are integrated bidirectionally with periodic wrapping in $x$ and clipping at the $y$ boundaries, and duplicate or very short curves are removed. The selected geodesic LCS are then compared with the operational FTLE-ridge masks using grid-neighborhood support fractions. This gives a variational/geodesic extraction that can be summarized through curve-length distributions, per-time curve counts, spatial occupancy, and agreement with the denser FTLE-ridge skeleton.

\subsection{Intersection, persistence, and scalar-gradient coupling}
Using time-matched forward and backward ridge masks, we compute forward ridge persistence maps, forward/backward intersection occupancy maps, and streamwise intersection profiles to identify where finite-time stretching repeatedly concentrates. Scalar-gradient coupling is then evaluated on ridges for temperature $\Temp$, mixture fraction $Z$, and a reaction intermediate HO$_2$. Ridge-conditioned gradient magnitudes and ridge-normal scalar-gradient projections for a scalar $\phi$ are computed as
\begin{equation}
g_{\phi,n}=\nabla \phi\cdot\hat{\mathbf{n}}, \qquad \hat{\mathbf{n}}=\frac{\nabla \sigma_T}{|\nabla \sigma_T|+\epsilon},
\end{equation}
where $\hat{\mathbf{n}}$ is taken from the FTLE gradient. This projection connects the local geometry of hyperbolic transport to scalar-gradient organization. We report either the signed projection or its absolute value as a geometric scalar-gradient measure. The scalar set spans thermal, mixing-state, and chemistry-conditioned responses. Temperature represents thermodynamic stratification, mixture fraction tracks molecular mixing, and HO$_2$ provides a representative intermediate from the 10-species kinetics mechanism.

\subsection{Deformation measures and sensitivity analysis}
In addition to the FTLE fields and geodesic-LCS curves, we use Cauchy--Green invariants ($\lambda_{\max}$, $\lambda_{\min}$, $\mathrm{tr}(\mathbf{C})$), shear-LCS metrics (IVD and LAVD), FTLE spectra along $x$, and forward/backward coherence-phase comparisons to place the ridge skeleton in the broader deformation field. Ridge-conditioned Cauchy statistics quantify how coherent structures sample deformation invariants differently from the full field.

\section{Results}
The results in this work are organized around the physical sequence from transport skeleton to scalar response. We begin with FTLE structure and ridge geometry, quantify forward/backward ridge interaction and deformation signatures, extract planar hyperbolic geodesic LCS from Cauchy--Green tensor fields, connect the transport structures to scalar transport, and then assess temporal coupling and robustness.

\subsection{FTLE structure and scalar overlays}
\begin{figure}[t]
  \centering
  \begin{subfigure}{0.95\linewidth}
    \includegraphics[width=\linewidth]{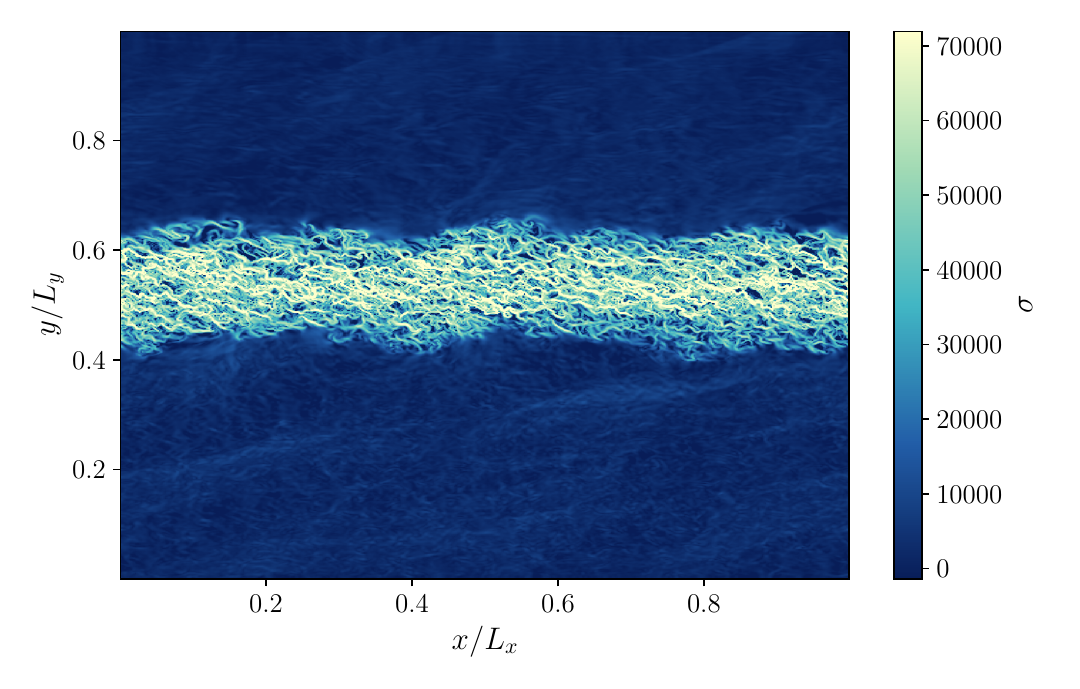}
    \caption{Forward FTLE with $\Temp$ contours}
  \end{subfigure}
  \par\medskip
  \begin{subfigure}{0.95\linewidth}
    \includegraphics[width=\linewidth]{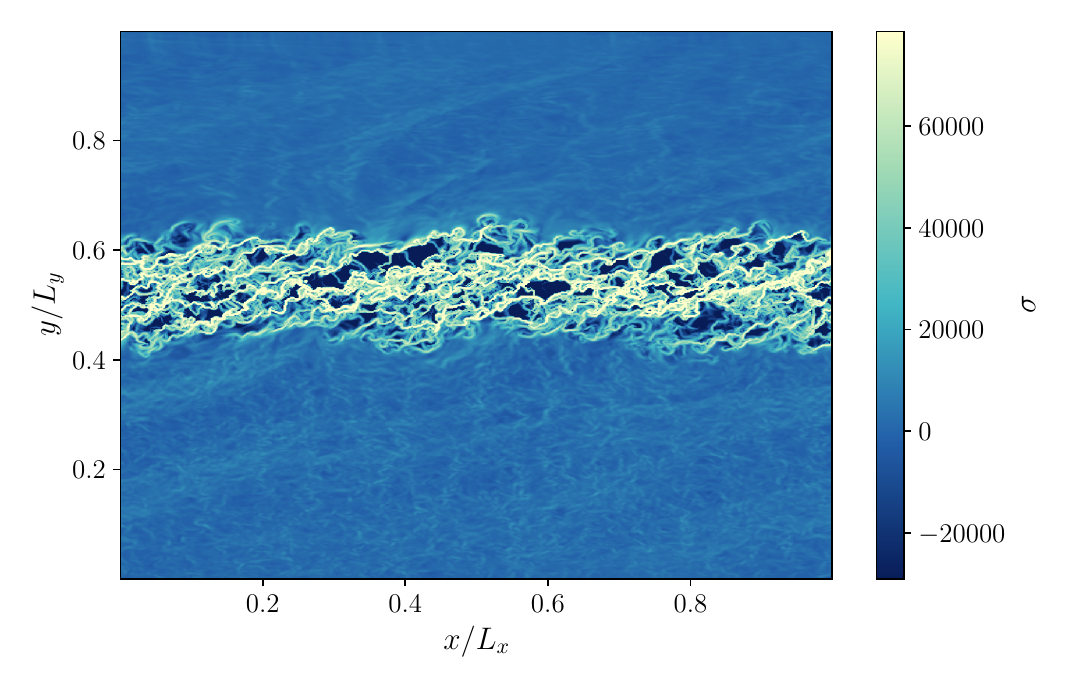}
    \caption{Backward FTLE}
  \end{subfigure}
  \caption{Representative FTLE fields in nondimensional coordinates, showing the attracting/repelling finite-time transport skeleton used throughout the planar analysis.}
  \label{fig:ftle_overlay}
\end{figure}

Figure~\ref{fig:ftle_overlay} establishes the central geometric pattern examined in this study. Thermochemical interfaces in the mid-plane slice are co-located with prominent hyperbolic FTLE ridges and have strongly nonuniform support across the background flow. Forward FTLE identifies repelling finite-time structures and backward FTLE identifies attracting finite-time structures; overlaid temperature isolines show steep thermal transitions clustered near the same regions. The expected forward/backward partition \citep{Shadden2005,Haller2015} sets up the ridge-conditioned scalar statistics reported later in the analysis.

\subsection{Integration-window sensitivity}
\begin{figure*}[t]
  \centering
  \begin{subfigure}{0.49\linewidth}
    \includegraphics[width=\linewidth]{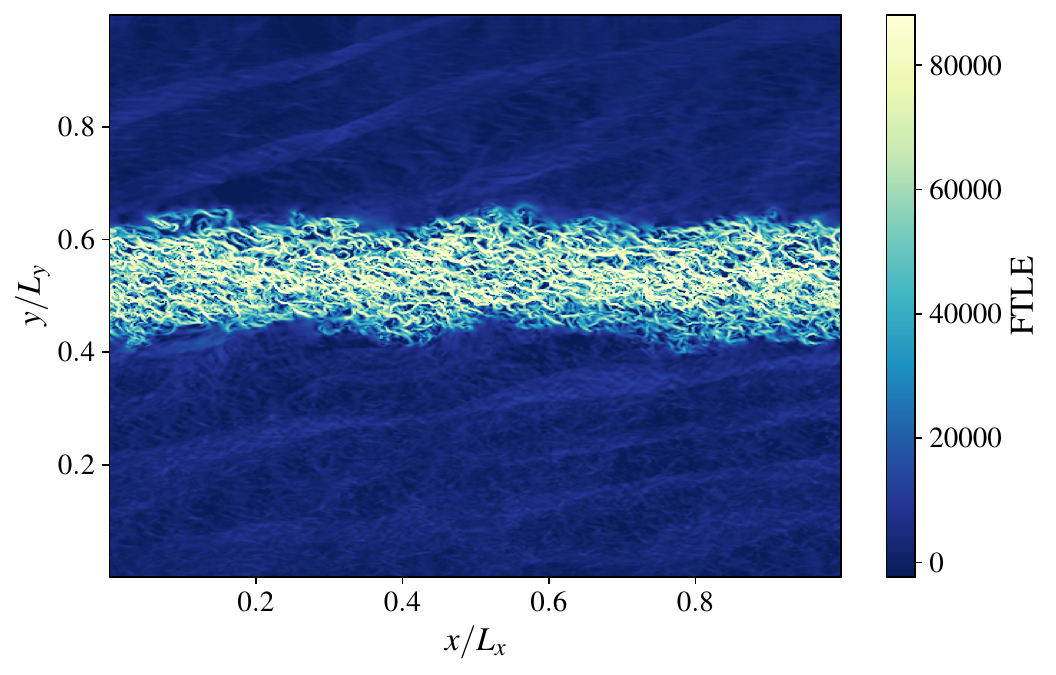}
    \caption{Forward, shorter horizon}
  \end{subfigure}
  \begin{subfigure}{0.49\linewidth}
    \includegraphics[width=\linewidth]{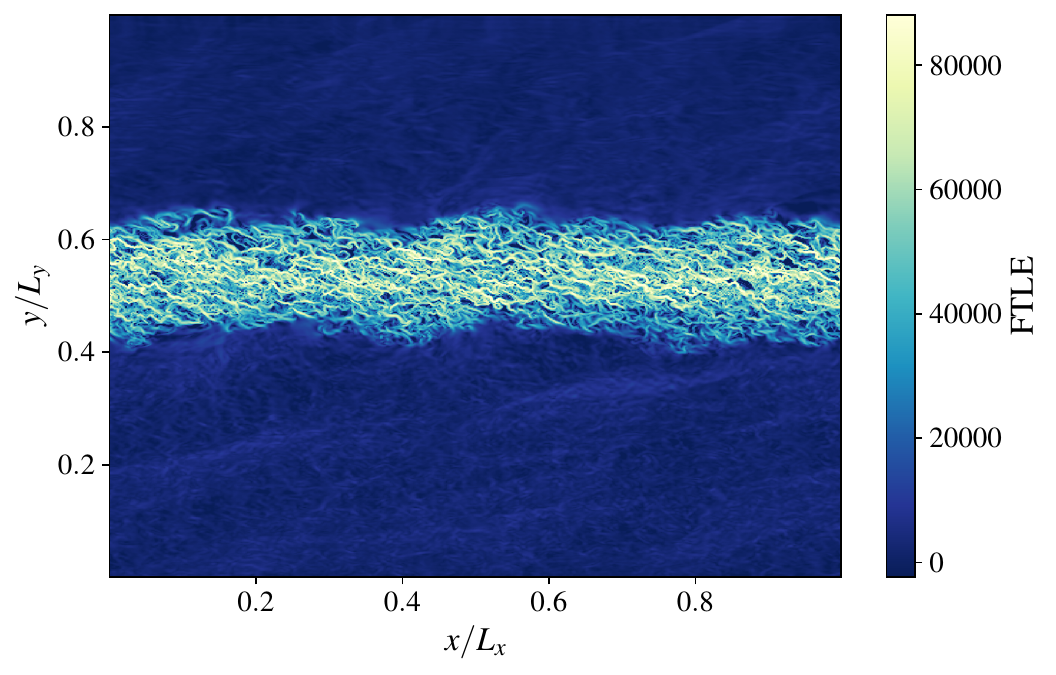}
    \caption{Forward, longer horizon}
  \end{subfigure}
  \begin{subfigure}{0.49\linewidth}
    \includegraphics[width=\linewidth]{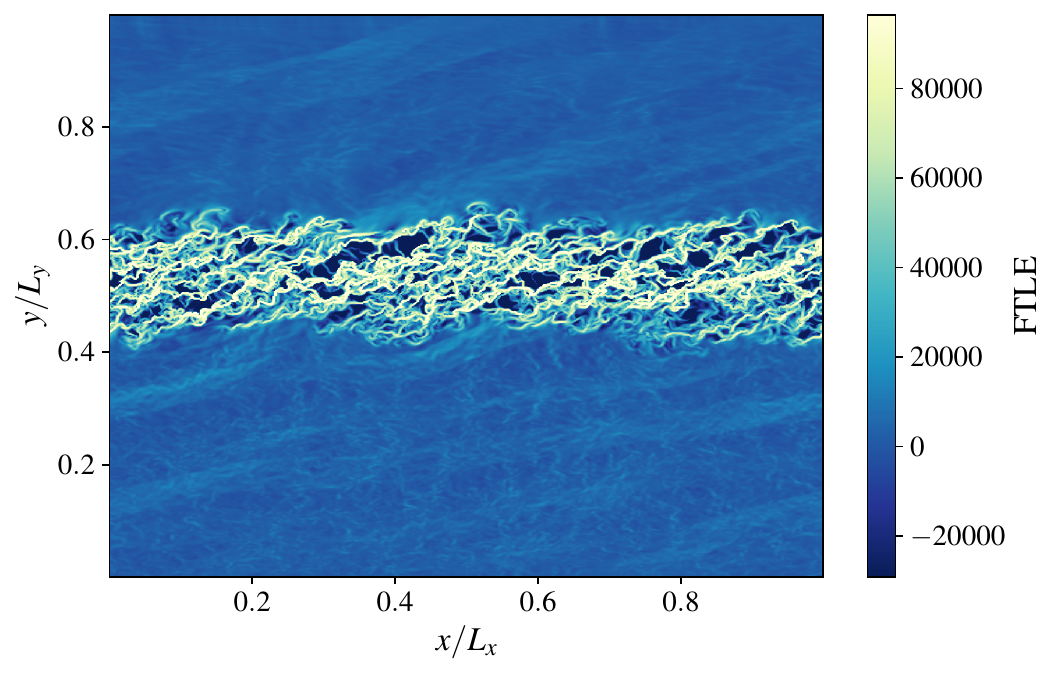}
    \caption{Backward, shorter horizon}
  \end{subfigure}
  \begin{subfigure}{0.49\linewidth}
    \includegraphics[width=\linewidth]{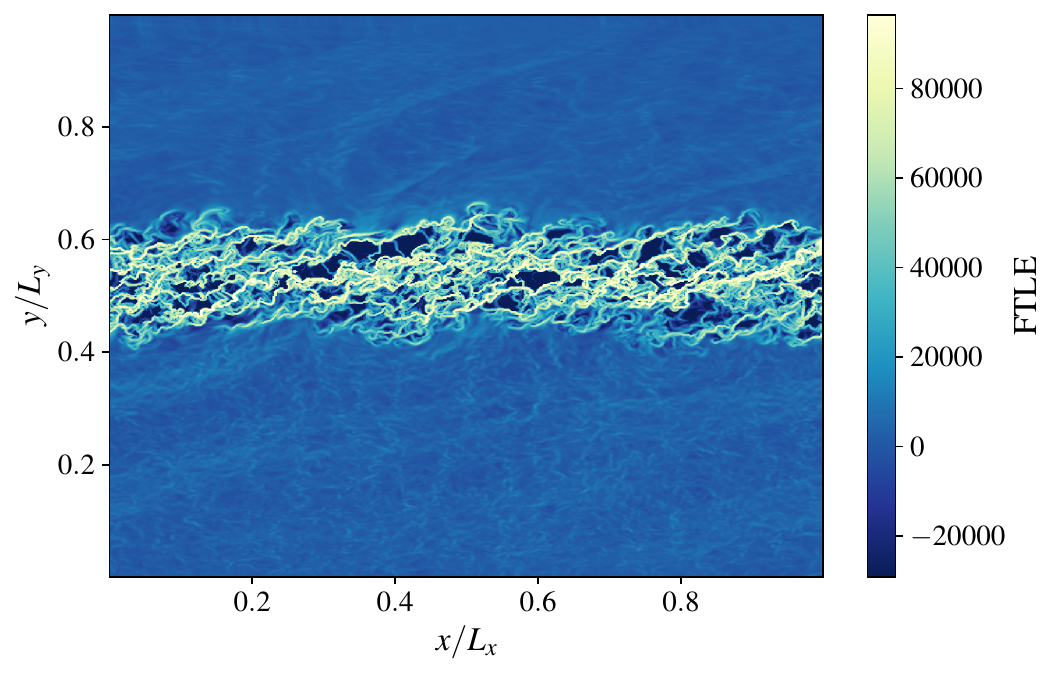}
    \caption{Backward, longer horizon}
  \end{subfigure}
  \caption{FTLE fields at a representative time for shorter and longer integration windows, showing persistence of the dominant finite-time transport skeleton under the tested window variation.}
  \label{fig:ftle_T_sensitivity}
\end{figure*}

The integration window for the FTLE ridges in Fig.~\ref{fig:ftle_overlay} must be long enough to resolve persistent finite-time structure and limit short-lived deformation noise. At the representative time shown in Fig.~\ref{fig:ftle_T_sensitivity}, the dominant FTLE-ridge skeleton remains topologically similar between the shorter and longer tested windows. The full-field correlations between the two windows are $0.945$ for forward FTLE and $0.923$ for backward FTLE, and $77\%$ and $83\%$ of the respective top-5\% FTLE grid cells are shared between the two windows. Increasing the horizon preserves the main ridge skeletons and damps weaker filamentary content, reducing the forward and backward 95th-percentile FTLE levels by about $15\%$ and $14\%$, respectively. The correlation and overlap measures together show that increasing the window mainly filters short-lived deformation noise and retains the persistent finite-time structures that organize planar transport, so the subsequent ridge statistics reflect robust finite-time structure under the tested windows \citep{AllshousePeacock2015,Gonzalez2016,LipinskiMohseni2010}.

\subsection{Ridge geometry and anisotropy trends}
\begin{figure}[t]
  \centering
  \begin{subfigure}{0.95\linewidth}
    \includegraphics[width=\linewidth]{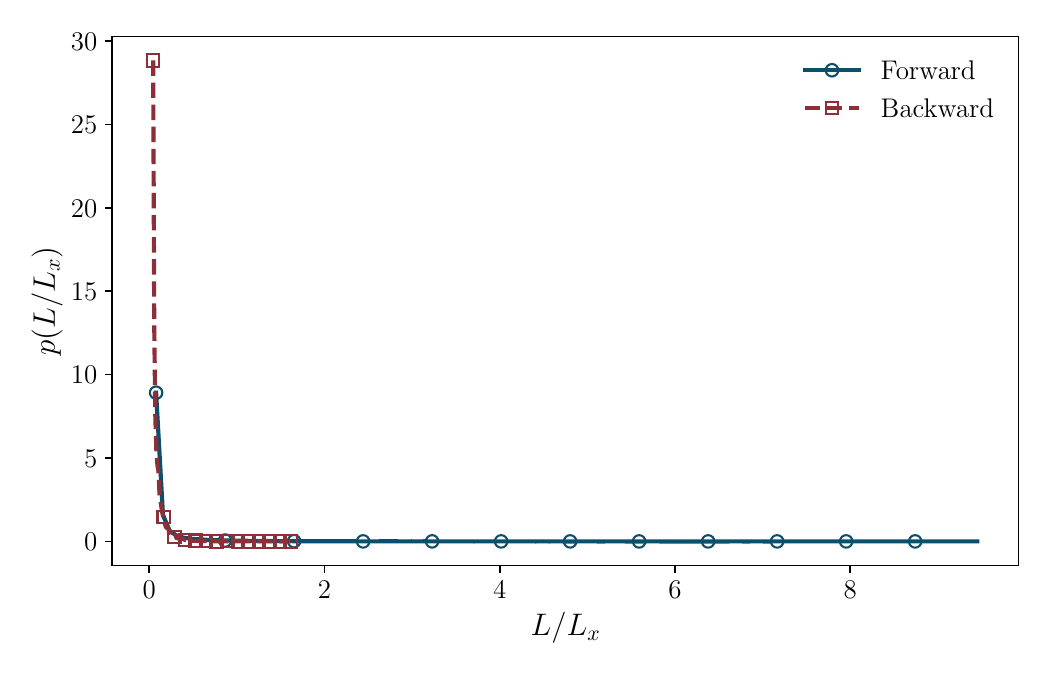}
    \caption{Ridge length PDF}
  \end{subfigure}
  \par\medskip
  \begin{subfigure}{0.95\linewidth}
    \includegraphics[width=\linewidth]{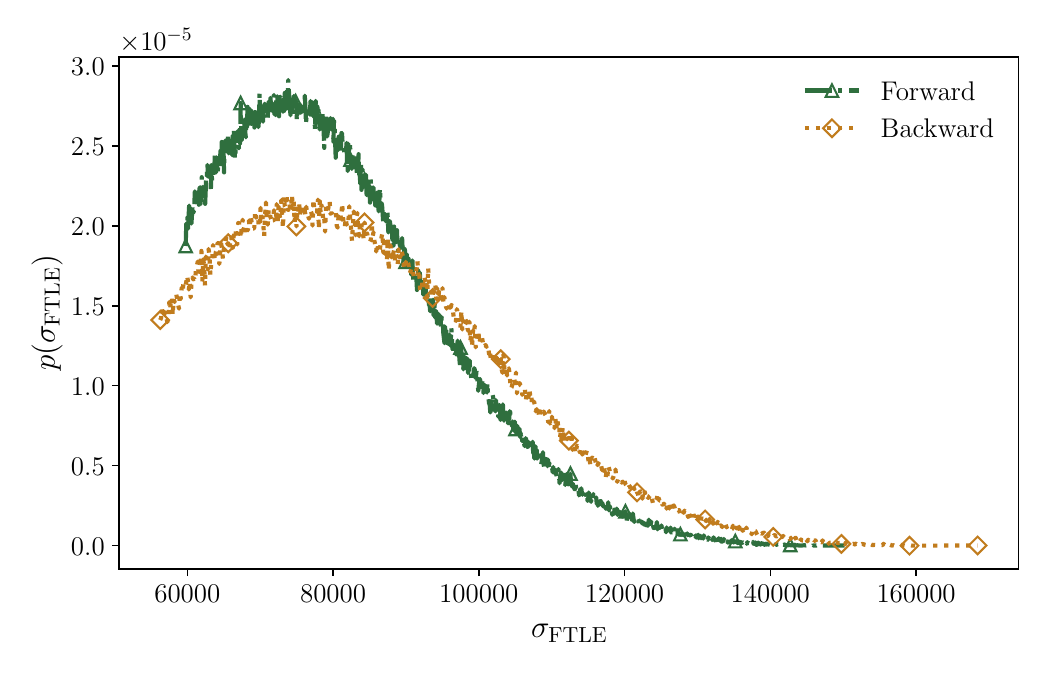}
    \caption{Ridge strength PDF}
  \end{subfigure}
  \par\medskip
  \begin{subfigure}{0.95\linewidth}
    \includegraphics[width=\linewidth]{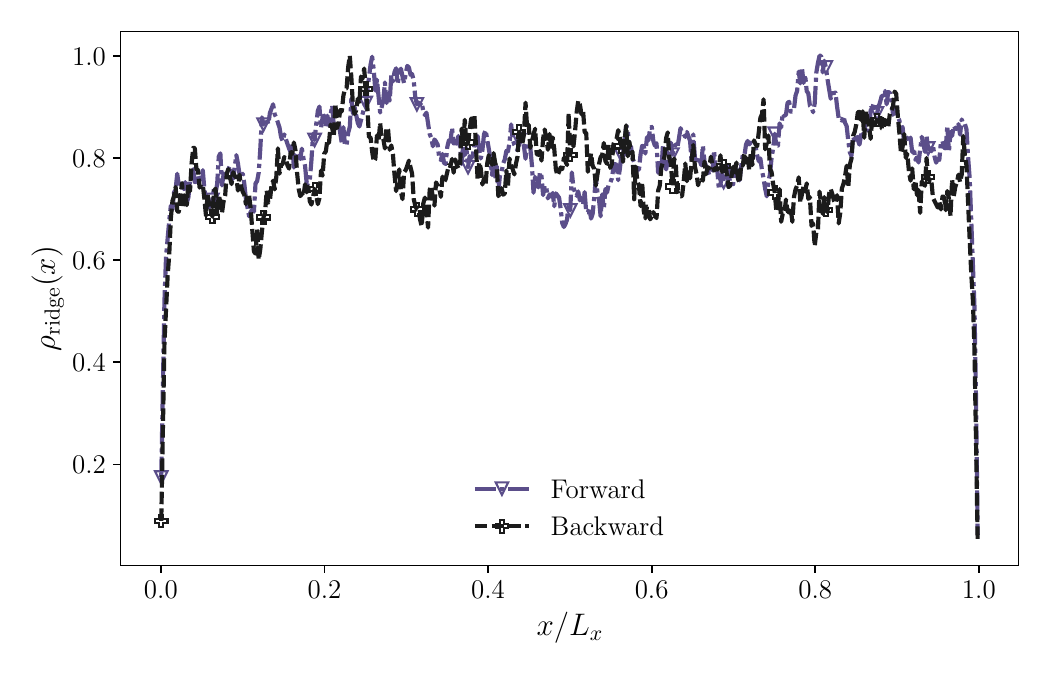}
    \caption{Streamwise ridge-occupancy profile}
  \end{subfigure}
  \caption{Combined forward/backward ridge statistics, summarizing geometric asymmetry in length, strength, and streamwise column-integrated occupancy.}
  \label{fig:ridge_combined}
\end{figure}

Table~\ref{tab:ridge_stats} and Fig.~\ref{fig:ridge_combined} quantify a clear directional asymmetry in ridge geometry. At the 95th-percentile level, forward ridges are nearly twice as long on average ($\langle L/L_x\rangle\approx1.8\times10^{-1}$) as backward ridges ($\approx9.4\times10^{-2}$). Backward ridges are modestly stronger in FTLE amplitude, with mean strength $\approx8.5\times10^{4}$ versus $\approx8.2\times10^{4}$. The streamwise concentration is also direction-dependent. The peak column-integrated ridge occupancy is higher for forward ridges ($\approx5.6$) than backward ridges ($\approx3.7$), implying broader forward occupancy of high-activity $x/L_x$ bands. These statistics indicate that geometric extent and local stretching intensity peak in different ridge families. Repelling structures occupy longer regions, and attracting structures carry slightly stronger local FTLE magnitude. The split has direct relevance for planar transport because long connected pathways and high local stretching play distinct roles in finite-time scalar redistribution. In a mechanical sense, the disparity is consistent with repelling and attracting ridge families acting as distinct transport organizers \citep{Shadden2005,Haller2001,Haller2015}.

\input{tables/ridge_stats_summary.tex}

The combined PDFs and streamwise occupancy profiles in Fig.~\ref{fig:ridge_combined} show that both the forward and backward families are strongly nonuniform in the streamwise direction, with their own persistent preferred zones. The nonuniformity implies spatially concentrated scalar-gradient enrichment near repeated ridge-occupancy peaks. Before any scalar-specific conditioning is applied, the ridge field already maps preferential transport corridors that provide a geometric basis for the enrichment and gradient-projection asymmetries reported later in this work.

\subsection{Threshold sensitivity and ridge tracking}
\begin{figure}[t]
  \centering
  \begin{subfigure}{0.95\linewidth}
    \includegraphics[width=\linewidth]{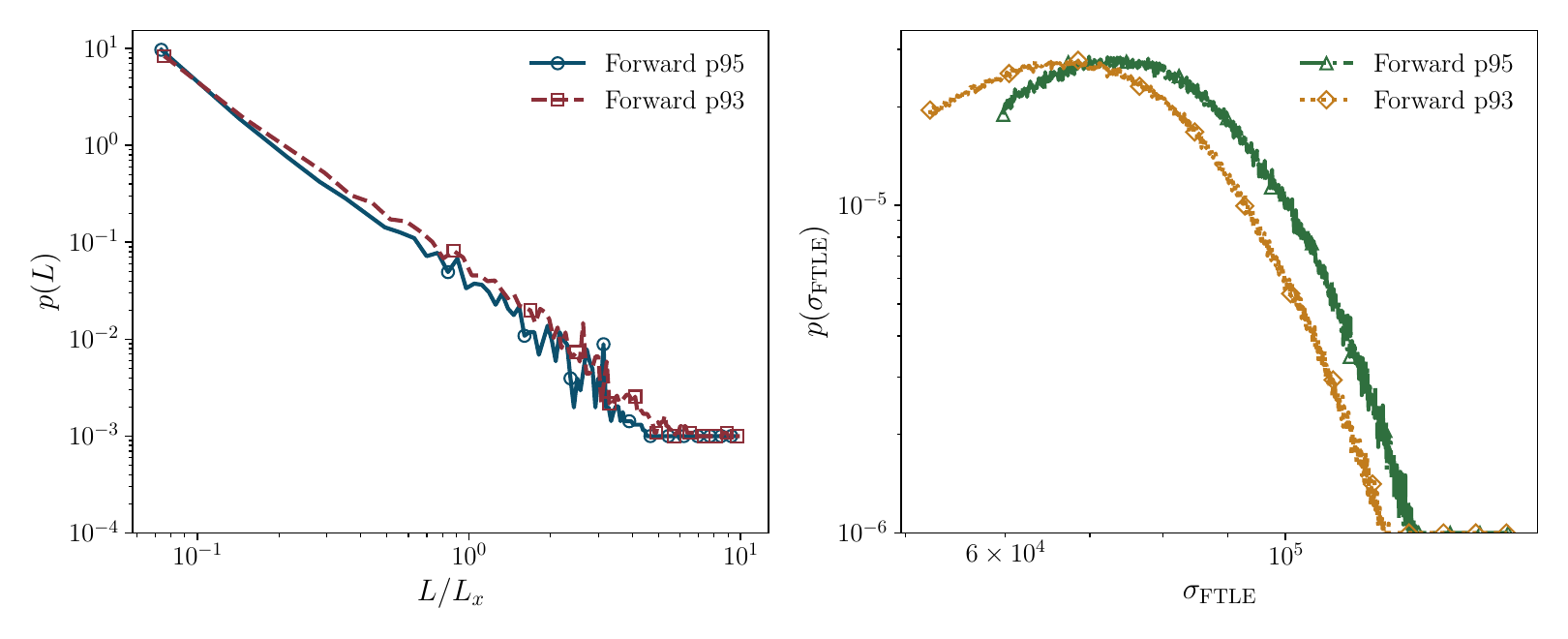}
    \caption{Forward}
  \end{subfigure}
  \begin{subfigure}{0.95\linewidth}
    \includegraphics[width=\linewidth]{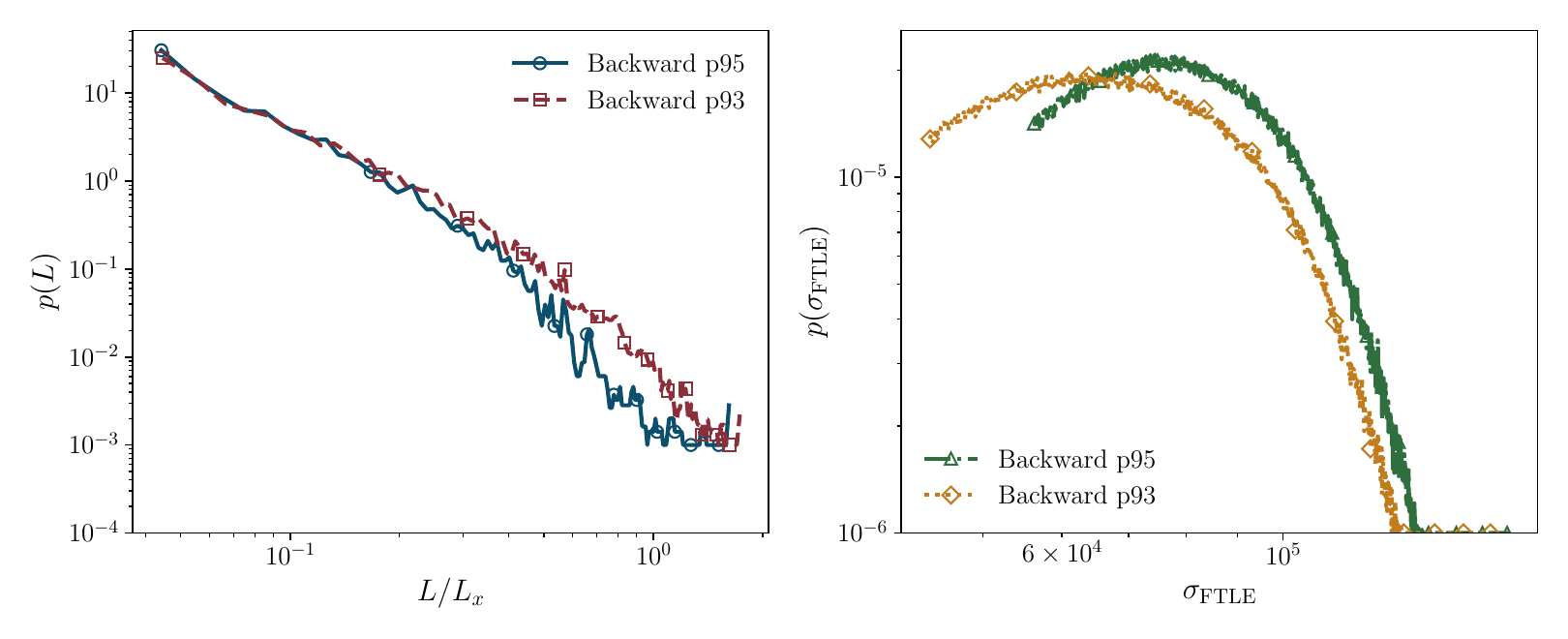}
    \caption{Backward}
  \end{subfigure}
  \caption{Ridge-statistics sensitivity to percentile threshold, showing consistent directional ordering under threshold relaxation.}
  \label{fig:ridge_sensitivity}
\end{figure}

The threshold-sensitivity results in Fig.~\ref{fig:ridge_sensitivity} show that lowering the cutoff from 95\% to 93\% increases ridge occupancy in both directions and preserves directional ordering. Forward mean ridge length shifts from $\approx1.8\times10^{-1}$ to $\approx2.3\times10^{-1}$, and backward shifts from $\approx9.4\times10^{-2}$ to $\approx1.1\times10^{-1}$. The corresponding mean changes are about $27\%$ and $20\%$, and the median lengths increase by about $20\%$ and $12.5\%$. The added structures have lower FTLE amplitude, with mean ridge strength dropping by about $9\%$ for forward ridges and $11\%$ for backward ridges. The growth in ridge counts and the drop in mean strength together indicate that the threshold-relaxed case mainly adds weaker secondary branches around an unchanged dominant FTLE-ridge skeleton and preserves the directional organization. Centroid-based component tracking (Fig.~\ref{fig:ridge_tracking}) yields $\approx1.4\times10^{4}$ forward and $\approx1.7\times10^{4}$ backward tracked components, with mean lifespans of $\approx1.9$ and $\approx2.2$ sampled frames, respectively. The lifespan distributions are dominated by single-frame components. About $3.5\%$ of forward ridges and $4.1\%$ of backward ridges persist beyond one sampled frame, and within that persistent subset the median lifespan is one sampled interval. The low-lifespan dominance indicates strong temporal intermittency of individual components. The persistence of population-level directional ordering indicates that the asymmetry is a recurrent ensemble property. Threshold variation and ridge tracking together show that the directional asymmetry persists across the tested extraction threshold and arises from ensemble organization beyond a few long-lived components.

\subsection{Forward/backward intersections and persistence}
\begin{figure}[t]
  \centering
  \begin{subfigure}{0.95\linewidth}
    \includegraphics[width=\linewidth]{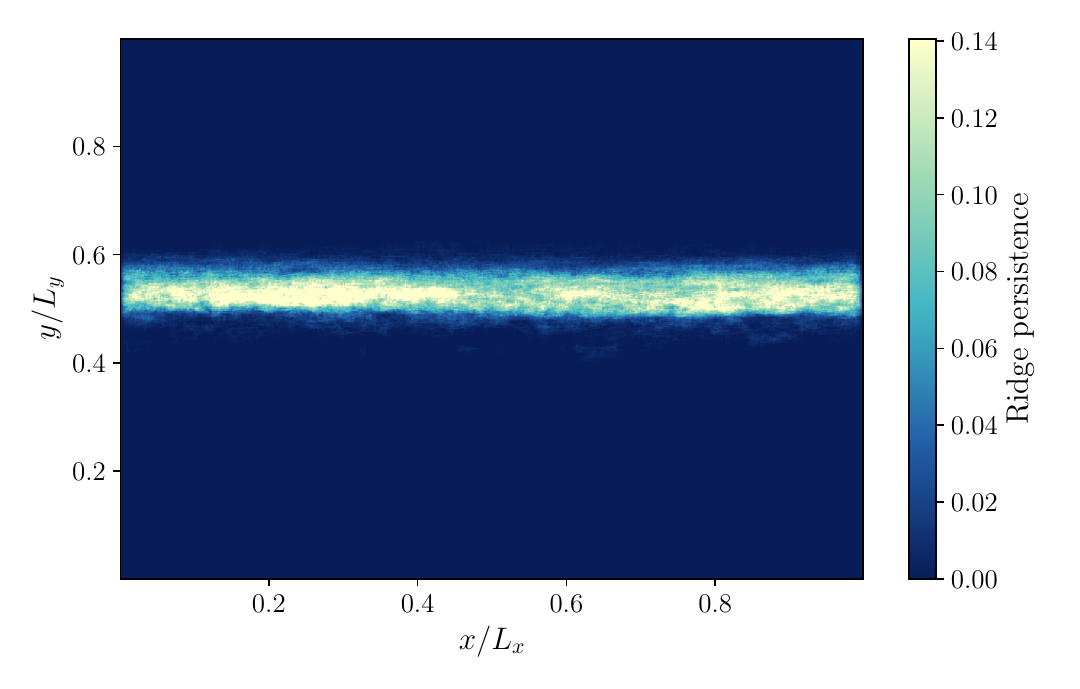}
    \caption{Forward ridge persistence}
  \end{subfigure}
  \par\medskip
  \begin{subfigure}{0.95\linewidth}
    \includegraphics[width=\linewidth]{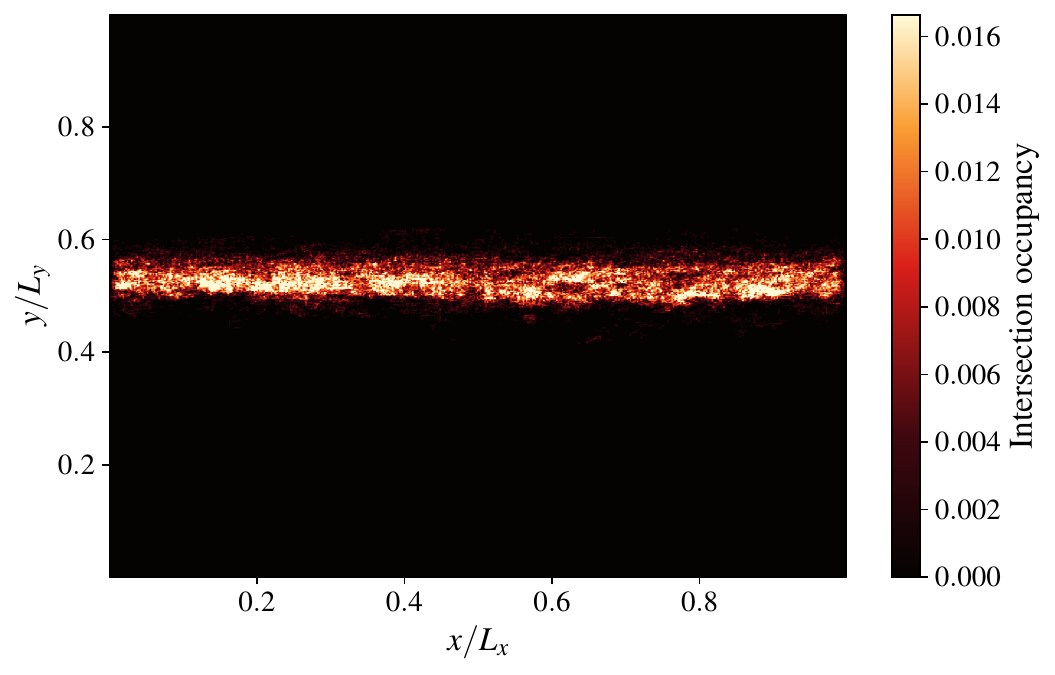}
    \caption{Forward/backward skeleton intersections}
  \end{subfigure}
  \caption{Persistence and intersection-occupancy maps, illustrating localized forward/backward ridge-interaction hotspots.}
  \label{fig:intersections}
\end{figure}

\begin{figure}[t]
  \centering
  \includegraphics[width=0.78\linewidth]{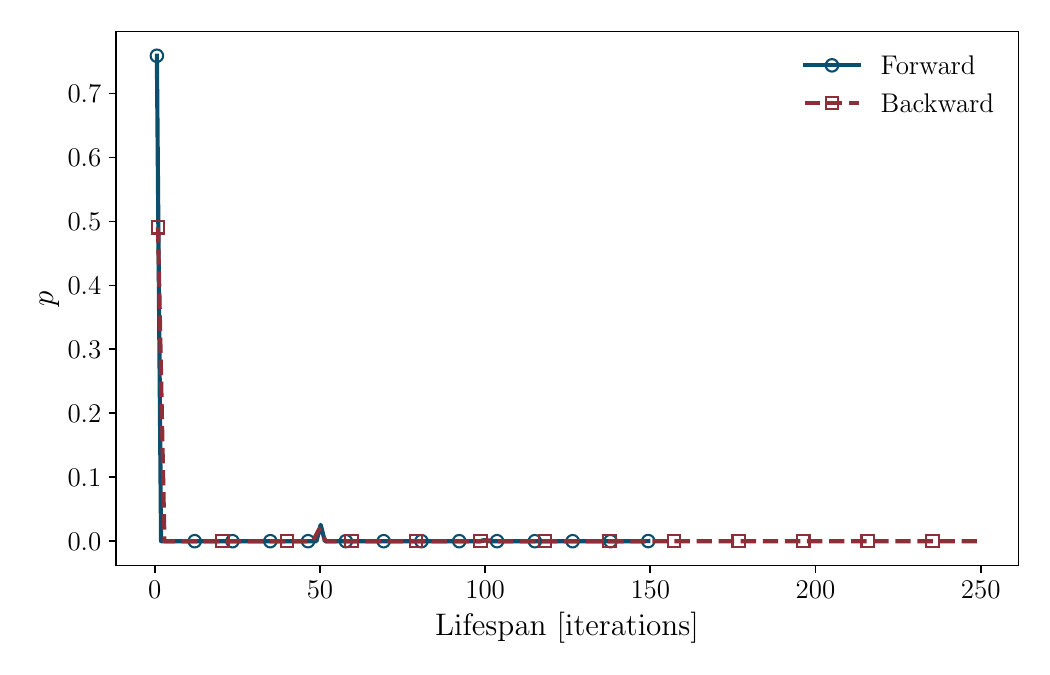}
  \caption{Forward/backward ridge lifespan distributions from centroid-based component tracking, indicating intermittent component-level dynamics.}
  \label{fig:ridge_tracking}
\end{figure}

Figure~\ref{fig:intersections} shows that forward/backward ridge interaction is highly localized to a small region confined inside the reacting shear layer. The mean skeleton-intersection occupancy is $\approx7.9\times10^{-4}$, and local maxima reach $\approx4.4\times10^{-2}$, so the strongest overlap locations are more than an order of magnitude larger than the domain-average background. About $3\%$ of the domain exceeds an intersection occupancy of $0.01$, and about $0.46\%$ exceeds $0.02$, showing that the overlap field is concentrated into a few hotspots with little broad-domain support. The streamwise intersection profile peaks near $x/L_x\approx0.26$, with a weaker secondary peak near $x/L_x\approx0.90$, indicating a primary interaction corridor in the mixing-layer core and a second one near the downstream edge. Forward persistence is similarly intermittent (mean $\approx9.3\times10^{-3}$, maximum $\approx2.3\times10^{-1}$), with about $4\%$ of the domain above a persistence level of $0.1$, so repeated ridge occupancy is spatially concentrated. The hotspot topology is consistent with the finite-time transport view that strongest stirring is organized around localized interaction zones \citep{Haller2015,Kelley2013,Mathur2007}. Along with the short component lifespans in Fig.~\ref{fig:ridge_tracking}, these maps indicate transient individual structures that repeatedly re-form in preferred corridors, yielding persistent hotspots despite intermittent component identity. Ridge overlap identifies the preferred recurrence locations of planar transport events.

\subsection{Cauchy--Green tensor, shear metrics, and spectral behavior}
\begin{figure*}[t]
  \centering
  \begin{subfigure}{0.49\linewidth}
    \includegraphics[width=\linewidth]{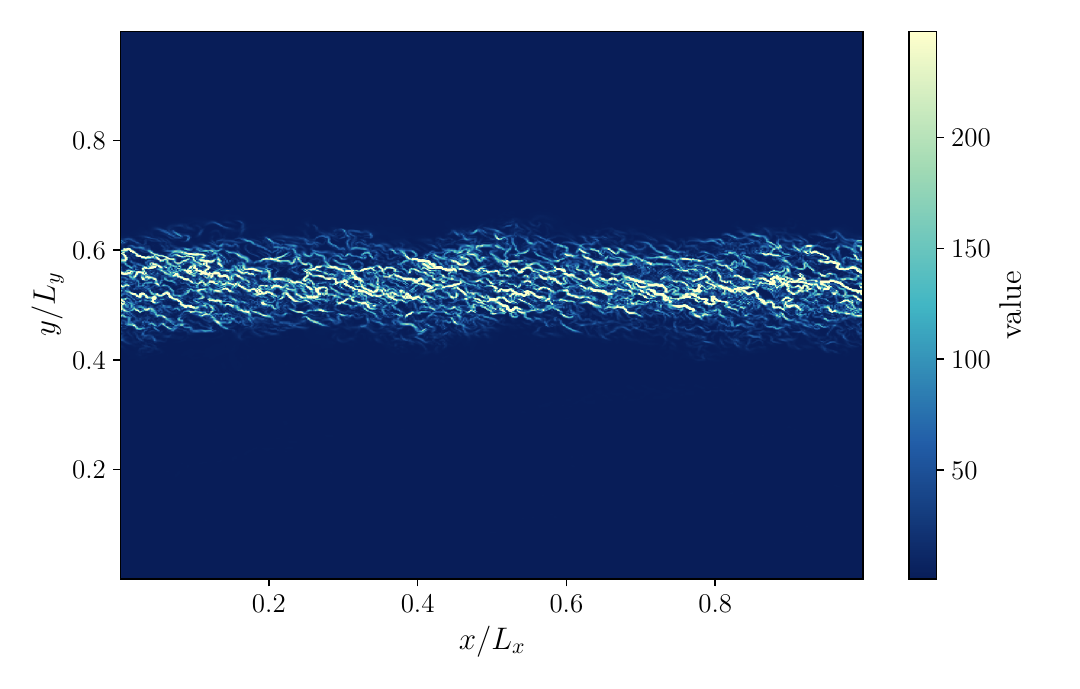}
    \caption{Forward $\lambda_{\max}$}
  \end{subfigure}
  \begin{subfigure}{0.49\linewidth}
    \includegraphics[width=\linewidth]{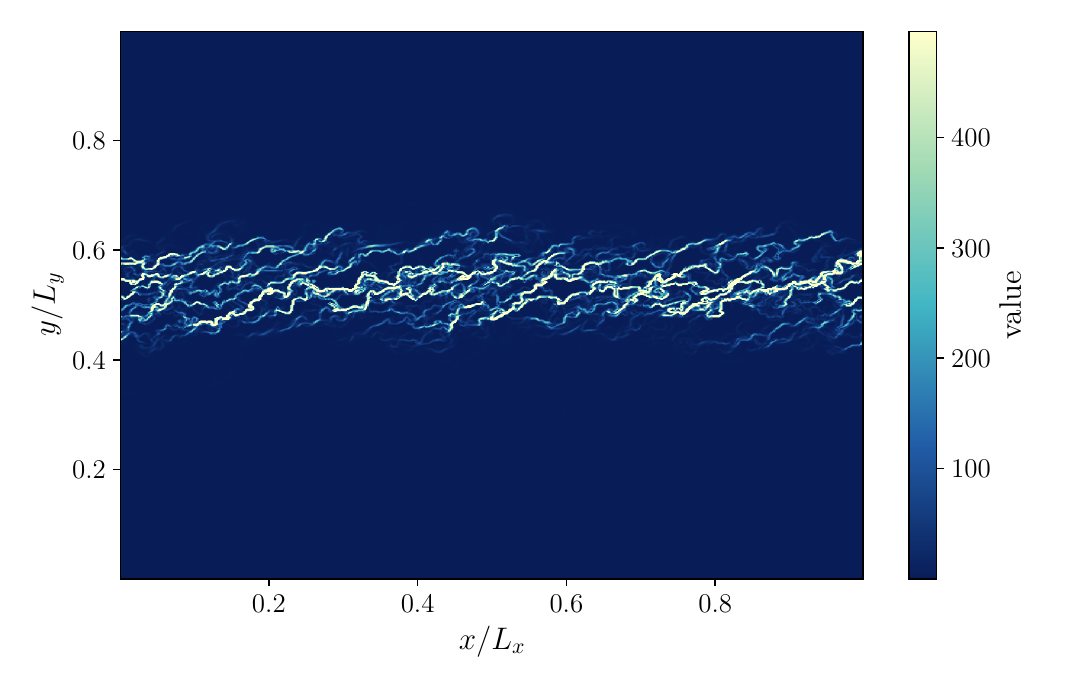}
    \caption{Backward $\lambda_{\max}$}
  \end{subfigure}
  \begin{subfigure}{0.49\linewidth}
    \includegraphics[width=\linewidth]{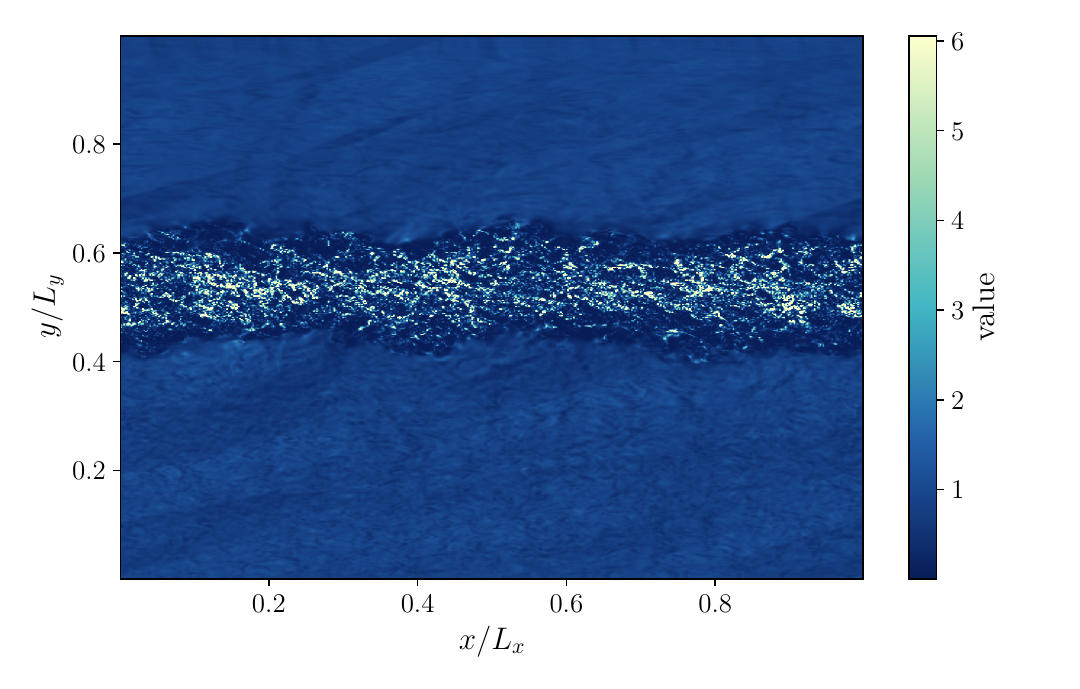}
    \caption{Forward $\lambda_{\min}$}
  \end{subfigure}
  \begin{subfigure}{0.49\linewidth}
    \includegraphics[width=\linewidth]{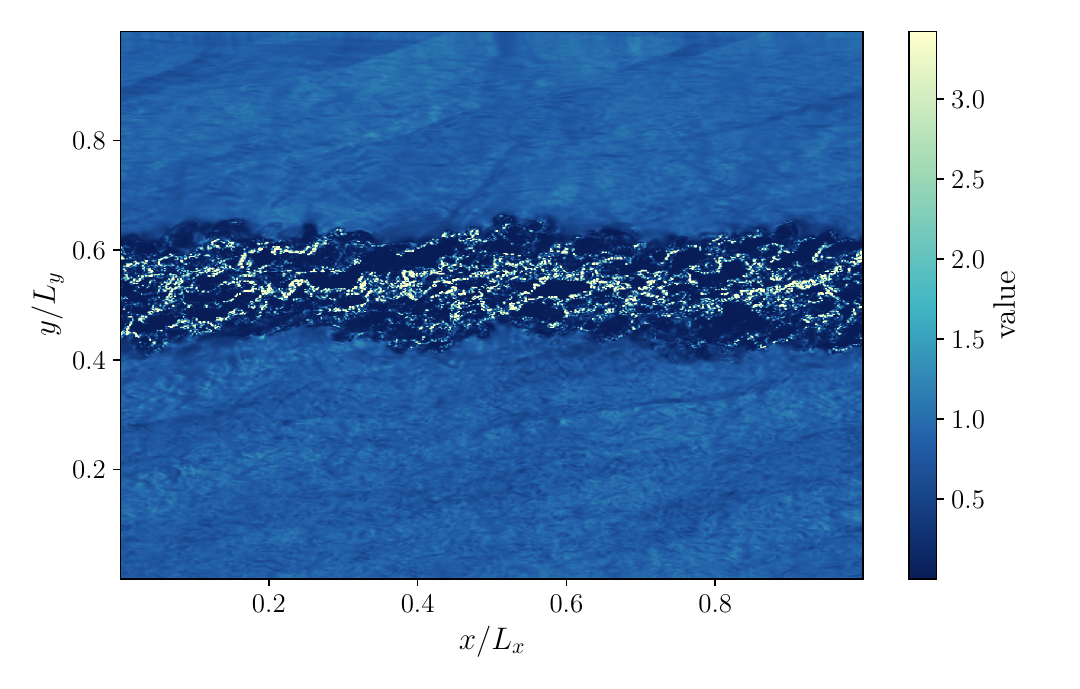}
    \caption{Backward $\lambda_{\min}$}
  \end{subfigure}
  \begin{subfigure}{0.49\linewidth}
    \includegraphics[width=\linewidth]{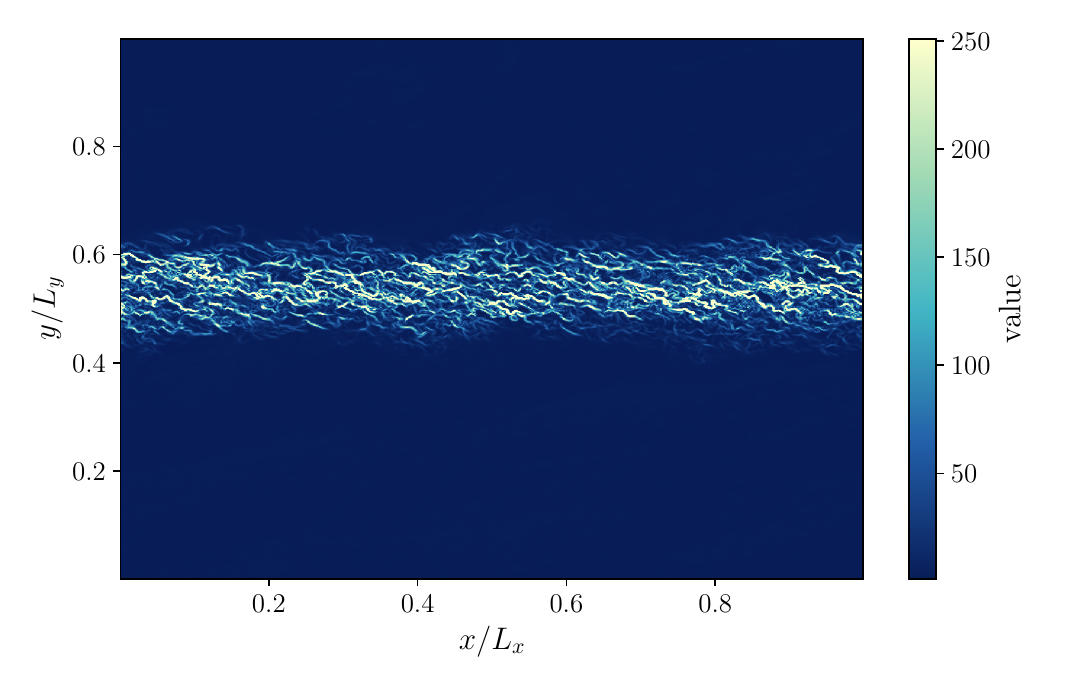}
    \caption{Forward $\mathrm{tr}(\mathbf{C})$}
  \end{subfigure}
  \begin{subfigure}{0.49\linewidth}
    \includegraphics[width=\linewidth]{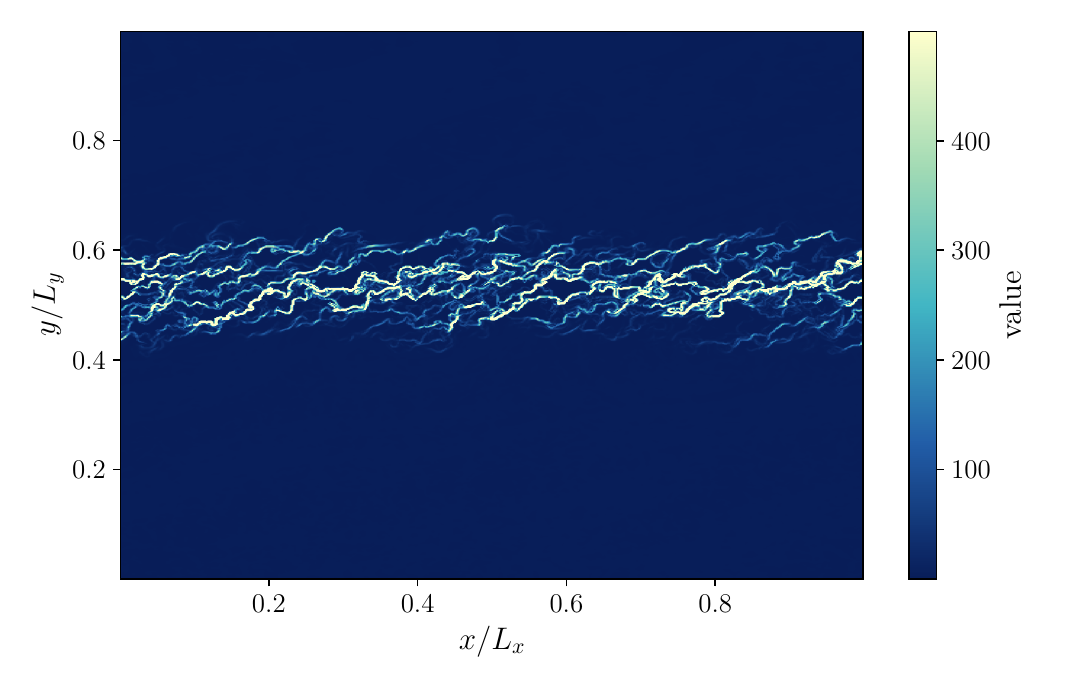}
    \caption{Backward $\mathrm{tr}(\mathbf{C})$}
  \end{subfigure}
  \caption{Cauchy--Green tensor invariant maps at representative time, showing deformation extrema sampled by the ridge field.}
  \label{fig:cauchy}
\end{figure*}

\begin{figure*}[t]
  \centering
  \begin{subfigure}{0.49\linewidth}
    \includegraphics[width=\linewidth]{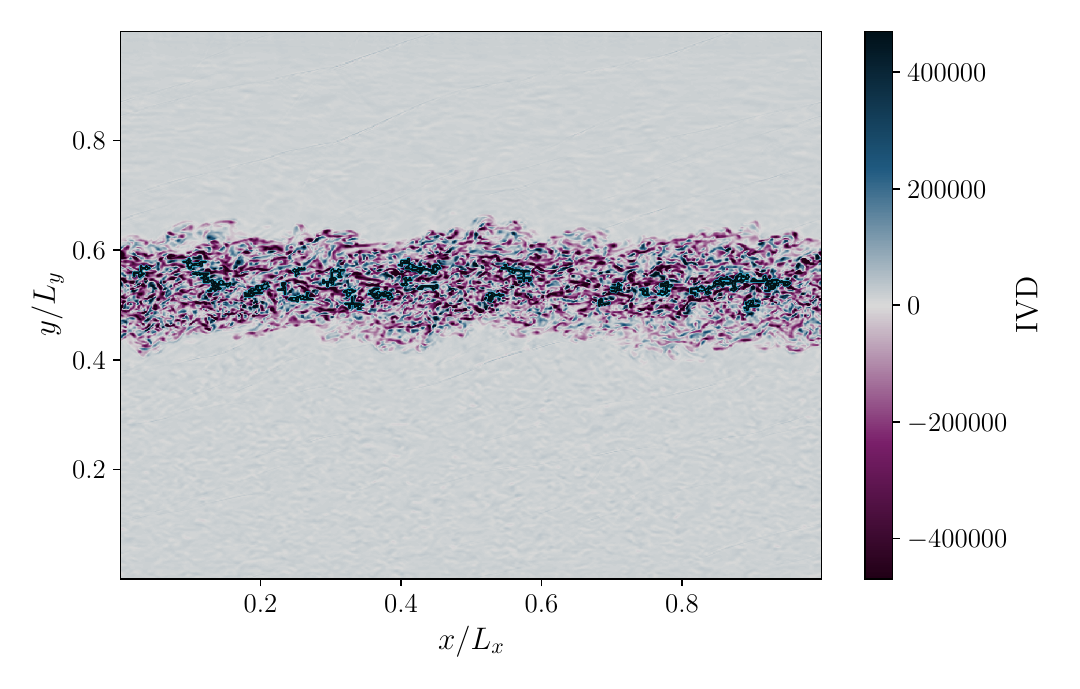}
    \caption{Forward IVD}
  \end{subfigure}
  \begin{subfigure}{0.49\linewidth}
    \includegraphics[width=\linewidth]{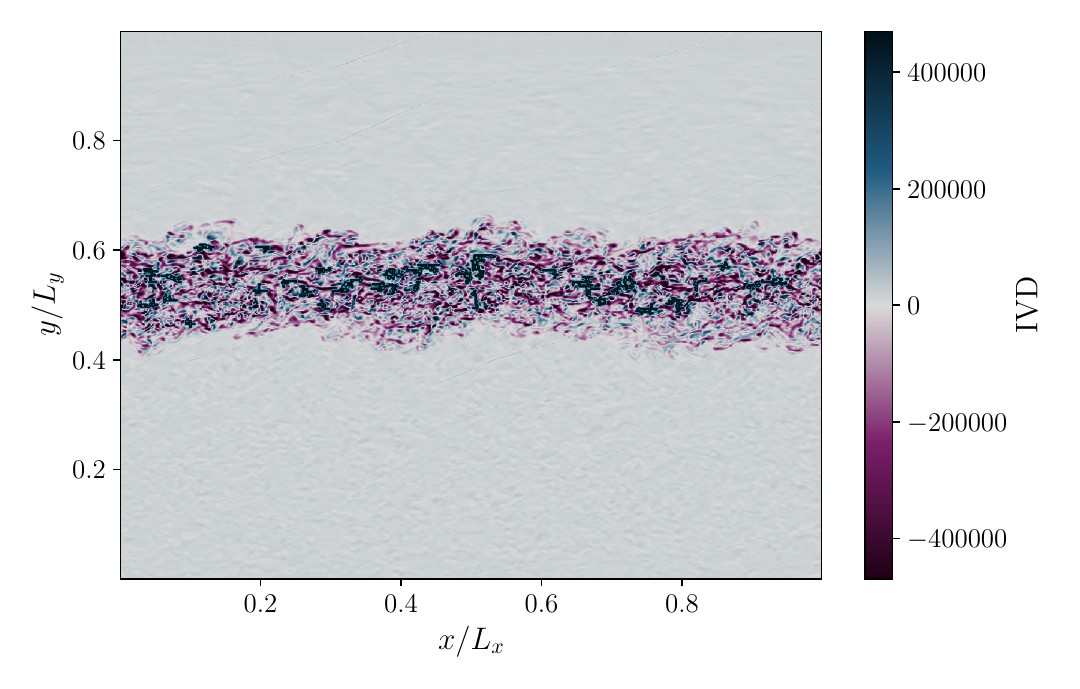}
    \caption{Backward IVD}
  \end{subfigure}
  \begin{subfigure}{0.49\linewidth}
    \includegraphics[width=\linewidth]{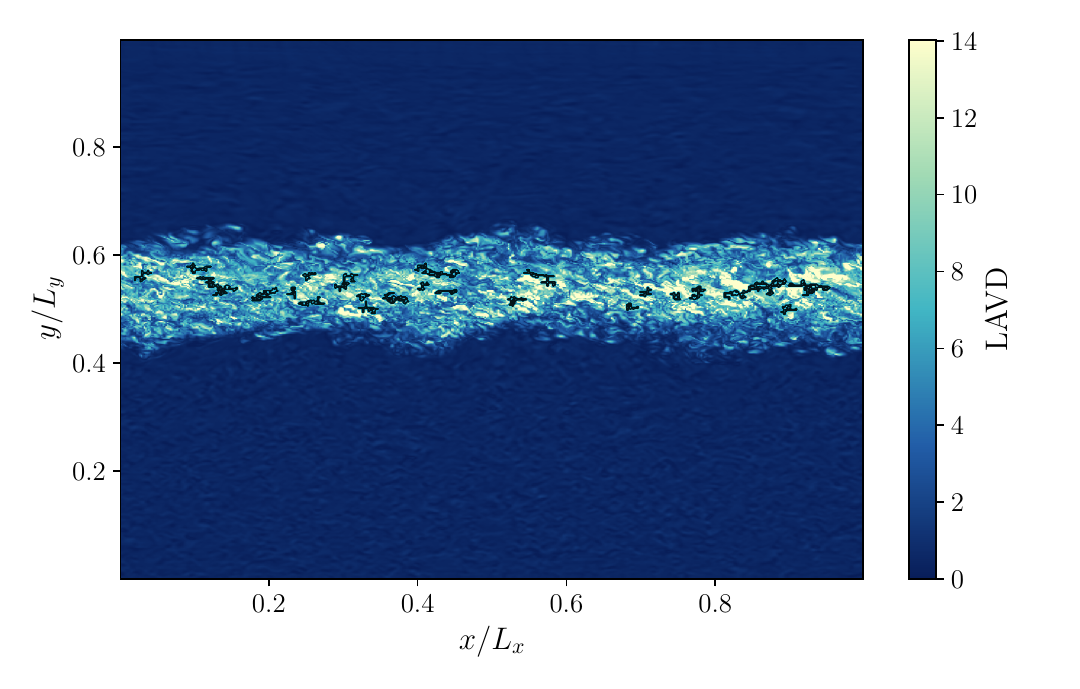}
    \caption{Forward LAVD}
  \end{subfigure}
  \begin{subfigure}{0.49\linewidth}
    \includegraphics[width=\linewidth]{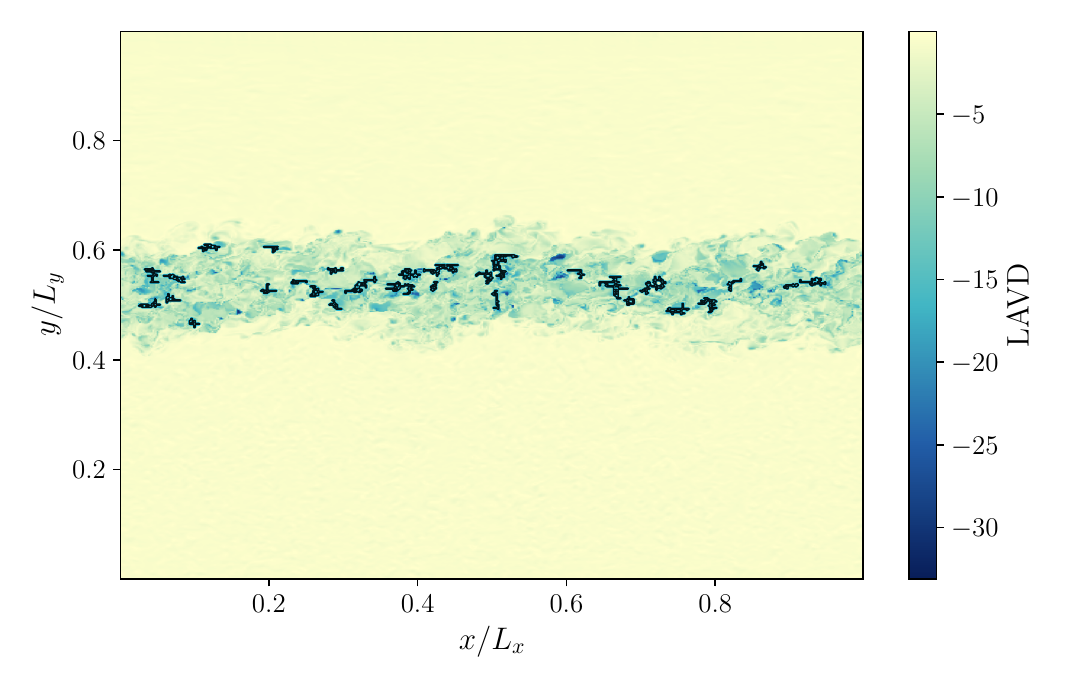}
    \caption{Backward LAVD}
  \end{subfigure}
  \caption{Shear-LCS metrics with ridge overlays, contrasting rotational coherence with hyperbolic transport structures.}
  \label{fig:shear_metrics}
\end{figure*}

Figure~\ref{fig:cauchy} shows that extracted ridges sample extreme finite-time deformation states characteristic of the shear layer. In particular, for the forward-time FTLE, the ridge-conditioned means are an order of magnitude larger than full-field means. The value of $\lambda_{\max}$ is $\approx2.4\times10^{2}$ on ridges versus $\approx1.3\times10^{1}$ over all points, and $\mathrm{tr}(\mathbf{C})$ is $\approx2.4\times10^{2}$ versus $\approx1.4\times10^{1}$. The separation shows that the ridge extraction preferentially samples high-strain regions. In parallel, the IVD/LAVD overlays in Fig.~\ref{fig:shear_metrics} show partial spatial overlap with hyperbolic ridges, indicating that rotational coherence and largely hyperbolic transport organization coexist as kinematically distinct structures \citep{Haller2001,Haller2015}.

\begin{figure}[t]
  \centering
  \begin{subfigure}{0.95\linewidth}
    \includegraphics[width=\linewidth]{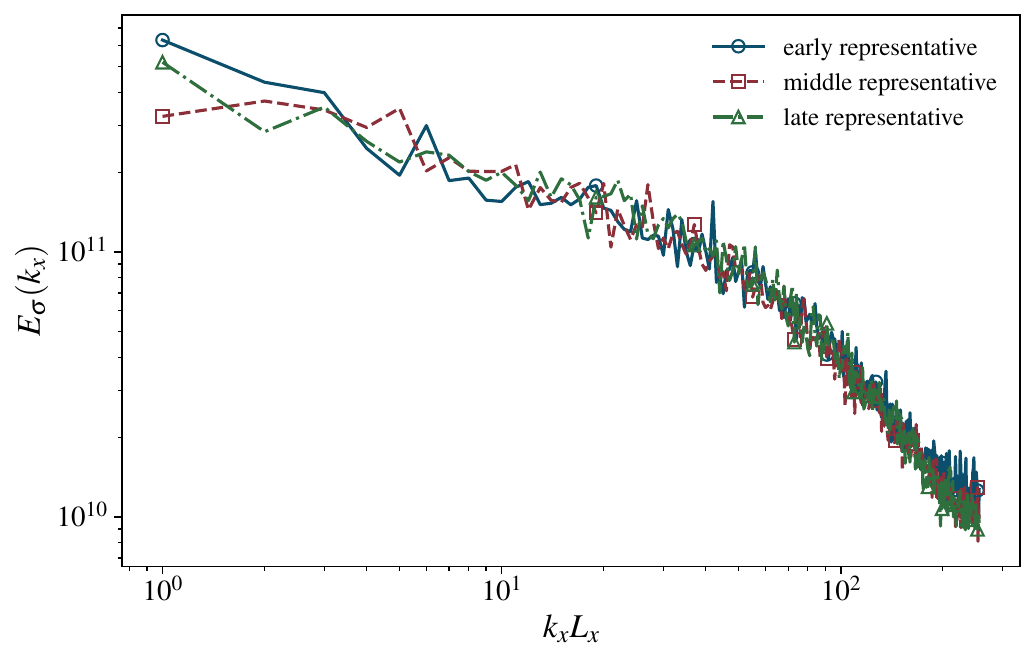}
    \caption{Forward FTLE spectra}
  \end{subfigure}
  \par\medskip
  \begin{subfigure}{0.95\linewidth}
    \includegraphics[width=\linewidth]{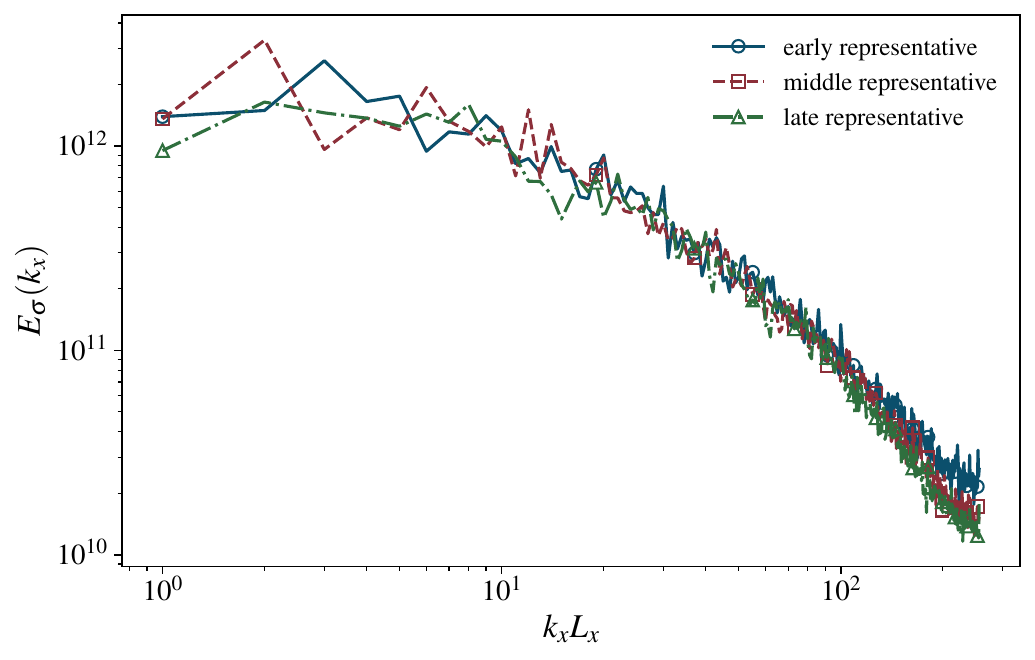}
    \caption{Backward FTLE spectra}
  \end{subfigure}
  \caption{FTLE streamwise spectra for three representative time instants, showing directional amplitude asymmetry across wavenumber bands.}
  \label{fig:ftle_spectra}
\end{figure}

The spectral view in Fig.~\ref{fig:ftle_spectra} shows that forward and backward FTLE fields occupy similar wavenumber support with systematically different amplitudes. Using low/mid/high $k_x$ partitions, forward mean power levels are $\approx(1.2\times10^{11},\,3.0\times10^{10},\,1.3\times10^{10})$, and backward levels are $\approx(4.5\times10^{11},\,6.1\times10^{10},\,2.2\times10^{10})$. The backward field carries higher streamwise FTLE spectral power across all bands, consistent with stronger attracting-structure concentration as observed earlier. Because this separation persists across low, mid, and high bands, it supports a broad directional asymmetry in deformation content. The spectra are consistent with the geometric result that forward and backward ridge families are related but quantitatively distinct transport families.

\subsection{Planar hyperbolic geodesic LCS}
\begin{figure*}[t]
  \centering
  \includegraphics[width=0.98\linewidth]{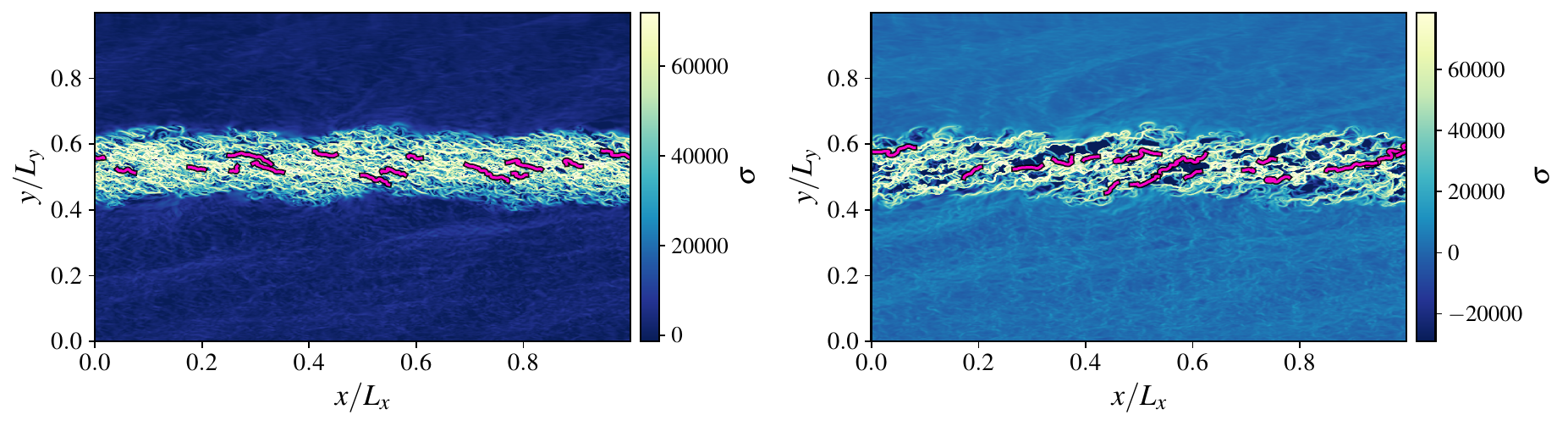}
  \caption{Planar hyperbolic geodesic LCS at a mid-window start time. Background shading is the FTLE field $\sigma$. Magenta curves are selected geodesic LCS; thin white contours are the operational FTLE-ridge masks. The left and right panels show the repelling and attracting families, respectively.}
  \label{fig:geodesic_lcs}
\end{figure*}

The geodesic extraction supplies the variational counterpart to the operational FTLE-ridge skeleton. Figure~\ref{fig:geodesic_lcs} shows a mid-window extraction in which the selected curves are strainlines of the minimum-eigenvalue eigenvector field seeded at high-$\lambda_{\max}$ normal maxima. The curves lie within the same high-strain shear-layer transport skeleton as the FTLE ridges and retain the material-curve subset satisfying the Cauchy--Green normal-maximum criterion.

\begin{figure}[t]
  \centering
  \includegraphics[width=0.98\linewidth]{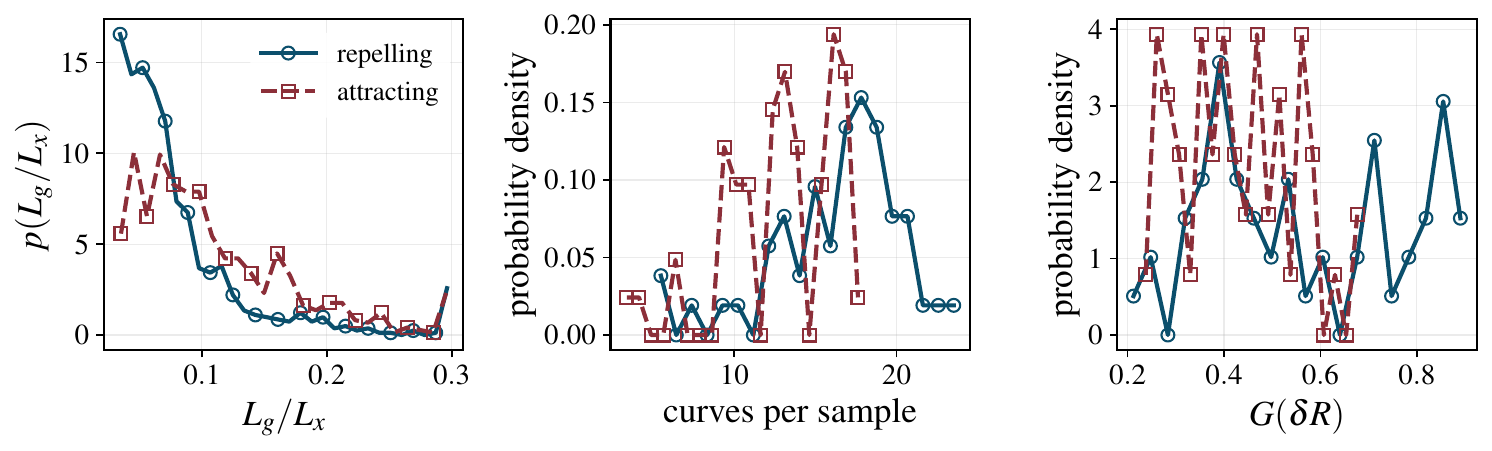}
  \caption{Time-window statistics of the planar geodesic-LCS extraction over 55 start times per family. The panels show pooled curve-length distributions, per-sample selected-curve counts, and geodesic support within the FTLE-ridge neighborhood.}
  \label{fig:geodesic_lcs_stats}
\end{figure}

The time-window statistics in Fig.~\ref{fig:geodesic_lcs_stats} and Table~\ref{tab:geodesic_lcs_timeseries} are computed from 55 reconstructed Cauchy--Green tensors per family spanning the synchronized interval. The extraction yields 904 repelling and 706 attracting geodesic curves. Repelling samples contain more curves per time on average ($16.44\pm4.10$ versus $12.84\pm3.46$). Attracting curves are longer, with median $L_g/L_x=0.092$ compared with $0.064$ for the repelling family. The mean geodesic-neighborhood agreement with the FTLE-ridge skeleton is $0.570$ for repelling curves and $0.427$ for attracting curves, and the reciprocal ridge-neighborhood support is nearly the same for both families ($\approx0.30$). The geodesic extraction selects a smaller set of material strainlines embedded within a broader operational ridge skeleton.

\begin{figure*}[t]
  \centering
  \includegraphics[width=0.98\linewidth]{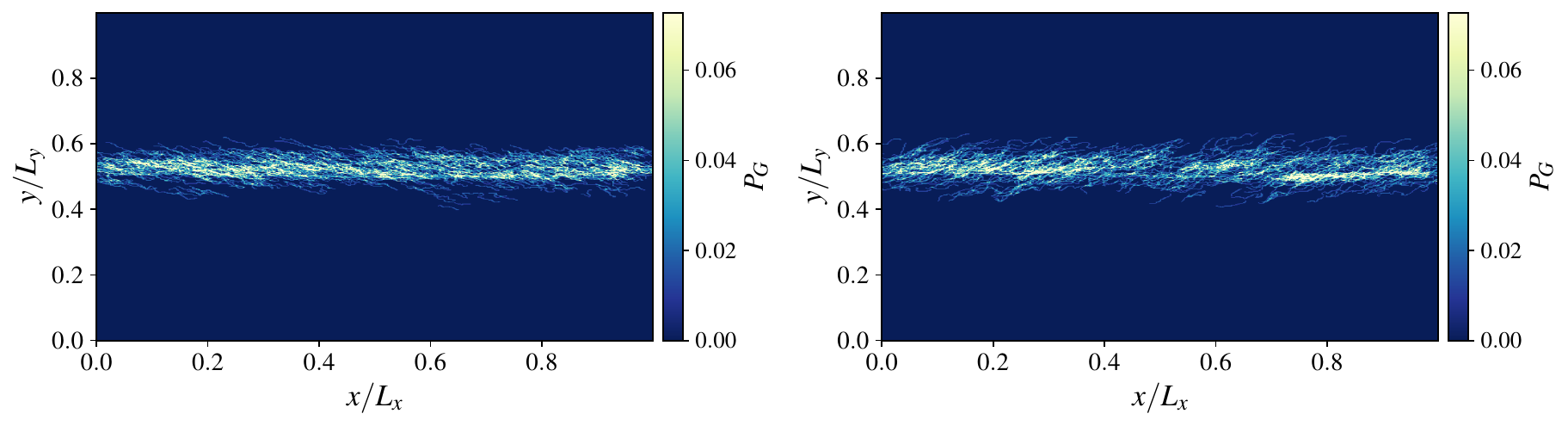}
  \caption{Spatial occupancy of selected planar geodesic LCS over 55 start times per family. The left and right panels show repelling and attracting geodesic occupancy, respectively, normalized by the number of sampled start times.}
  \label{fig:geodesic_lcs_occupancy}
\end{figure*}

The occupancy maps in Fig.~\ref{fig:geodesic_lcs_occupancy} show that geodesic LCS repeatedly localize inside the same mid-layer band as the FTLE ridges. Peak geodesic occupancy reaches $\approx0.16$ for the repelling family and $\approx0.20$ for the attracting family over the 55 sampled start times, and the domain-mean occupancy remains $\approx2.6\times10^{-3}$. The extracted curves form a sparse, recurrent material-curve population concentrated in the reacting shear layer, as expected for a variational geodesic subset of the full finite-time transport skeleton.

\begin{figure}[t]
  \centering
  \includegraphics[width=0.98\linewidth]{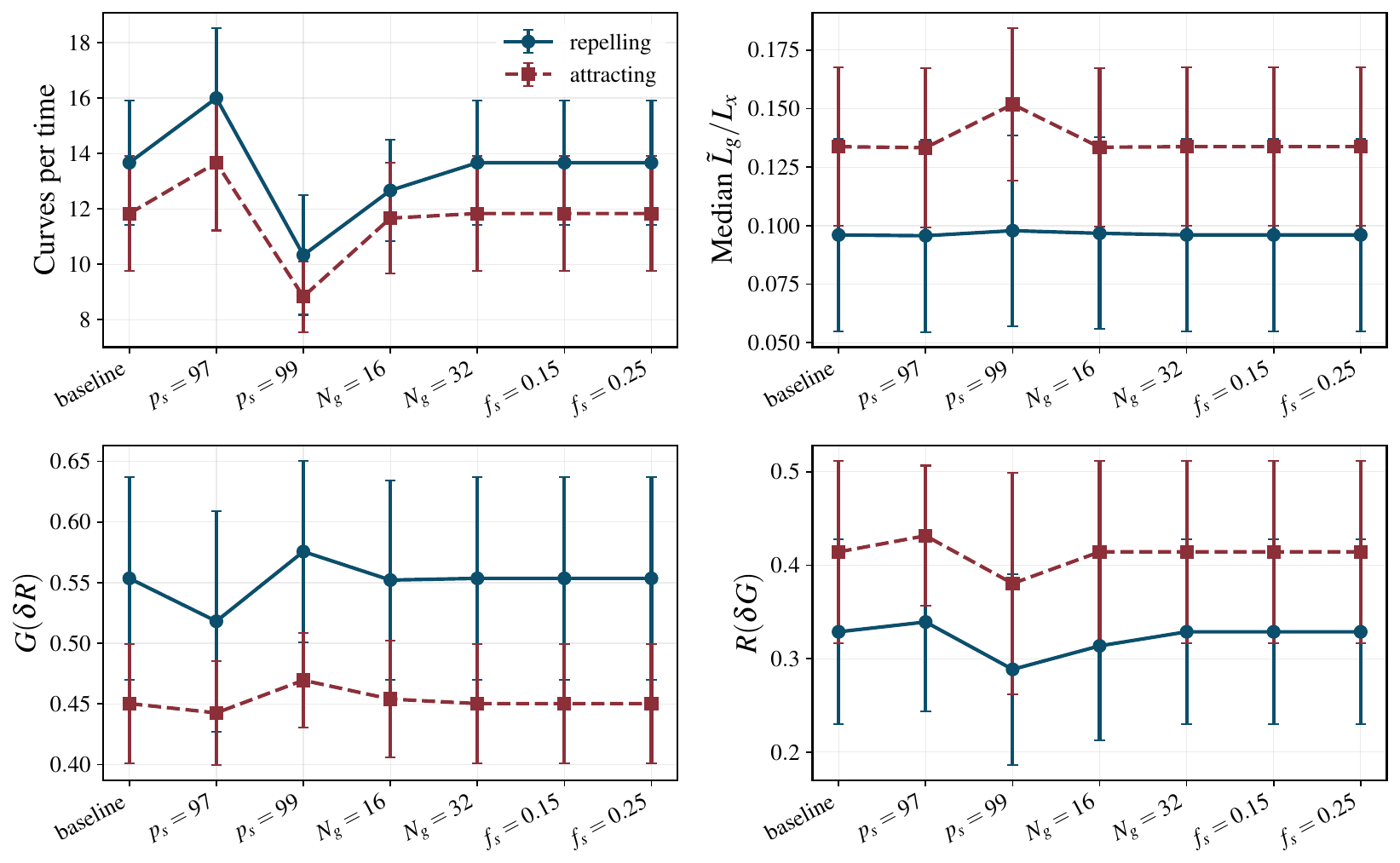}
  \caption{One-factor geodesic-extraction sensitivity over six matched start times per family, varying seed percentile $p_s$, retained-curve cap $N_g$, and minimum support fraction $f_s$ around the baseline. Error bars show standard error of the time mean over the matched subset.}
  \label{fig:geodesic_lcs_sensitivity}
\end{figure}

The geodesic extraction was also rerun on a matched six-start-time subset to test sensitivity to the three choices that most directly affect curve selection, seed percentile $p_s$, retained-curve cap $N_g$, and minimum normal-maximum support fraction $f_s$ (Fig.~\ref{fig:geodesic_lcs_sensitivity}, Table~\ref{tab:geodesic_lcs_sensitivity}). Changing $p_s$ from 97 to 99 alters curve counts as expected, and the median lengths and grid-neighborhood agreement fractions remain in the same range as the baseline. The stricter seed percentile increases selectivity and preserves the extracted family. Repelling $G(\delta R)$ remains between $0.518$ and $0.576$, and attracting $G(\delta R)$ remains between $0.443$ and $0.469$. The curve cap has a minor effect over this subset because most samples contain fewer than the baseline cap of 24 retained curves. The tested support-fraction range also leaves the summary unchanged, indicating that selected curves already satisfy the normal-maximum support criterion with margin. The sensitivity cases cover selected extraction settings and show that the geodesic result persists across seed-percentile, curve-cap, and support-threshold variations.

\input{tables/geodesic_lcs_timeseries_summary.tex}

\input{tables/geodesic_lcs_sensitivity.tex}

\subsection{Scalar coupling with ridges}
\begin{figure}[t]
  \centering
  \begin{subfigure}{0.95\linewidth}
    \includegraphics[width=\linewidth]{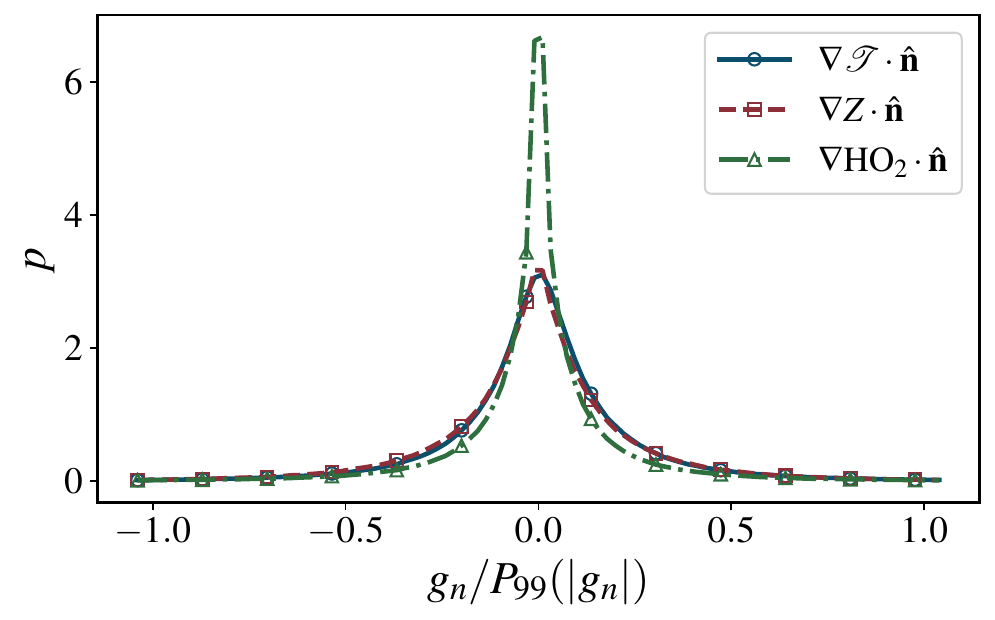}
    \caption{Forward normalized ridge projection PDFs}
  \end{subfigure}
  \par\medskip
  \begin{subfigure}{0.95\linewidth}
    \includegraphics[width=\linewidth]{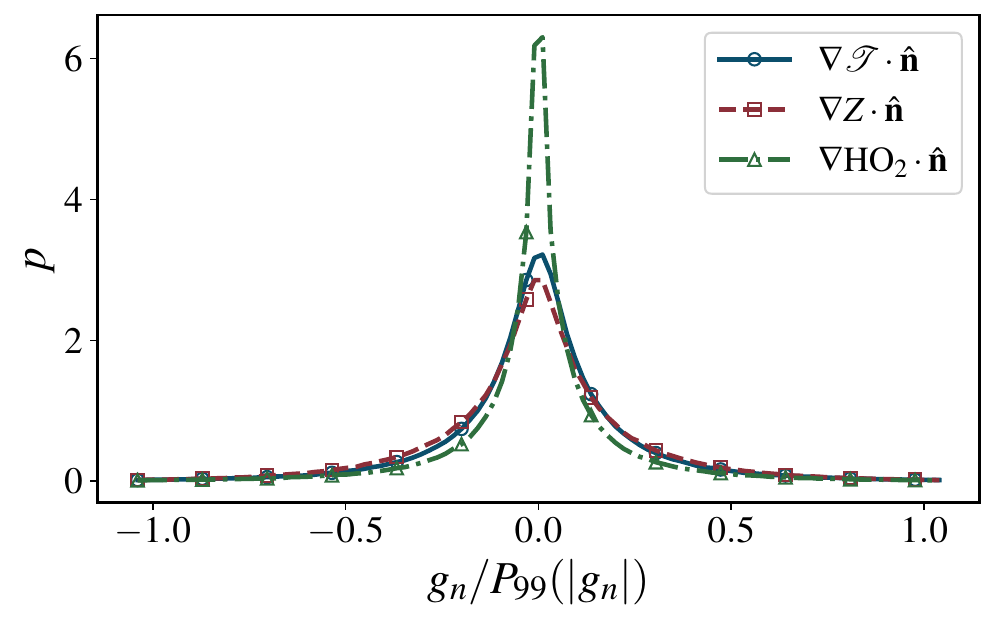}
    \caption{Backward normalized ridge projection PDFs}
  \end{subfigure}
  \par\medskip
  \begin{subfigure}{0.95\linewidth}
    \includegraphics[width=\linewidth]{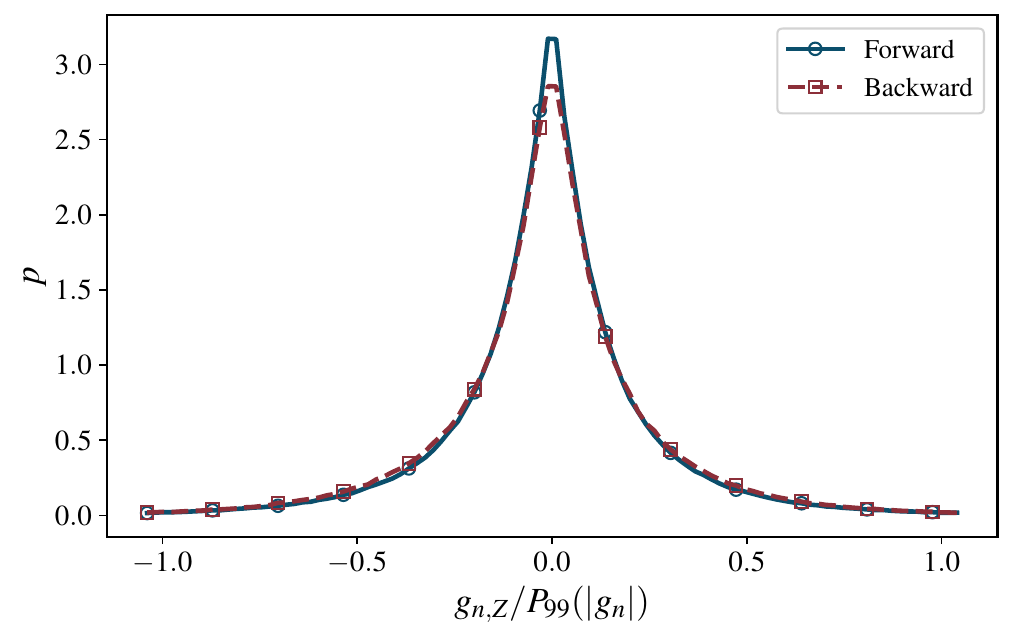}
    \caption{Combined $Z$ ridge-projection statistics}
  \end{subfigure}
  \caption{Scalar coupling and ridge-normal gradient-projection behavior, linking FTLE-ridge geometry to species/thermal transport selectivity.}
  \label{fig:scalar_coupling}
\end{figure}

Figure~\ref{fig:scalar_coupling} shows full-time ridge-normal projection PDFs on forward and backward ridges, linking the FTLE-ridge geometry to thermochemical scalar organization with directional (forward/backward) selectivity. These pooled statistics use all 541 synchronized ridge-intersection times. Table~\ref{tab:scalar_coupling_full_time} reports the corresponding ridge-conditioned scalar-gradient and absolute projection ratios. The full-time gradient enrichment factors (ridge mean/all-point mean) are $\approx6.34$ and $\approx6.54$ for $\Temp$ (forward/backward), $\approx9.53$ and $\approx9.48$ for $Z$, and $\approx4.09$ and $\approx5.85$ for HO$_2$. The larger enrichment in $Z$ relative to the other scalars indicates that ridge geometry most strongly co-locates with mixture-fraction gradient localization; thermodynamic and intermediate-species fields show similar weaker amplification. Ridge-normal gradient-projection PDFs show asymmetry between attracting and repelling families, indicating that transport-skeleton directionality is associated with both where gradients are large and how scalar gradients project along FTLE-normal directions \citep{Sampath2016,Frohlich2008,Eisma2021,VanderwelTavoularis2016}. The enriched HO$_2$ response remains below $Z$, consistent with an intermediate-species field whose gradients reflect both transport organization and finite-rate reaction-zone occupancy \citep{DopazoCifuentes2016,ChakrabortyDopazo2024}. The ordering $Z>\Temp\gtrsim\mathrm{HO_2}$ for forward ridges and $Z>\Temp>\mathrm{HO_2}$ for backward ridges indicates that coherent-structure conditioning is strongest for mixing-state gradients, with chemistry represented through conditional intermediate-species gradient response.

\input{tables/scalar_coupling_full_timeseries.tex}

\subsection{Null-model separation of ridge-conditioned scalar enrichment}
The enrichment factors in Fig.~\ref{fig:scalar_coupling} establish strong ridge conditioning. Separating coherent-structure selectivity from shear-layer localization requires a time- and cross-stream-stratified null model over the same 541 synchronized times. For each time, direction, scalar, and $y$ row, the null preserves the number of ridge grid cells in that row and randomizes their streamwise positions. The construction controls both temporal evolution and shear-layer-normal localization before forming the null ratio
\begin{equation}
R = \frac{\langle |\nabla\phi| \rangle_{\mathrm{subset}}}{\langle |\nabla\phi| \rangle_{\mathrm{all}}}.
\end{equation}
Under this null, $R$ reflects the scalar-gradient level obtained by a random mask with the same time-varying cross-stream support as the extracted ridges. We compare the observed ridge-conditioned ratio $R_{\mathrm{obs}}$ against the stratified null envelope and use the residual ratio $R_{\mathrm{obs}}/R_{\mathrm{strat}}$ as the main separation measure.

\begin{figure}[t]
  \centering
  \includegraphics[width=0.92\linewidth]{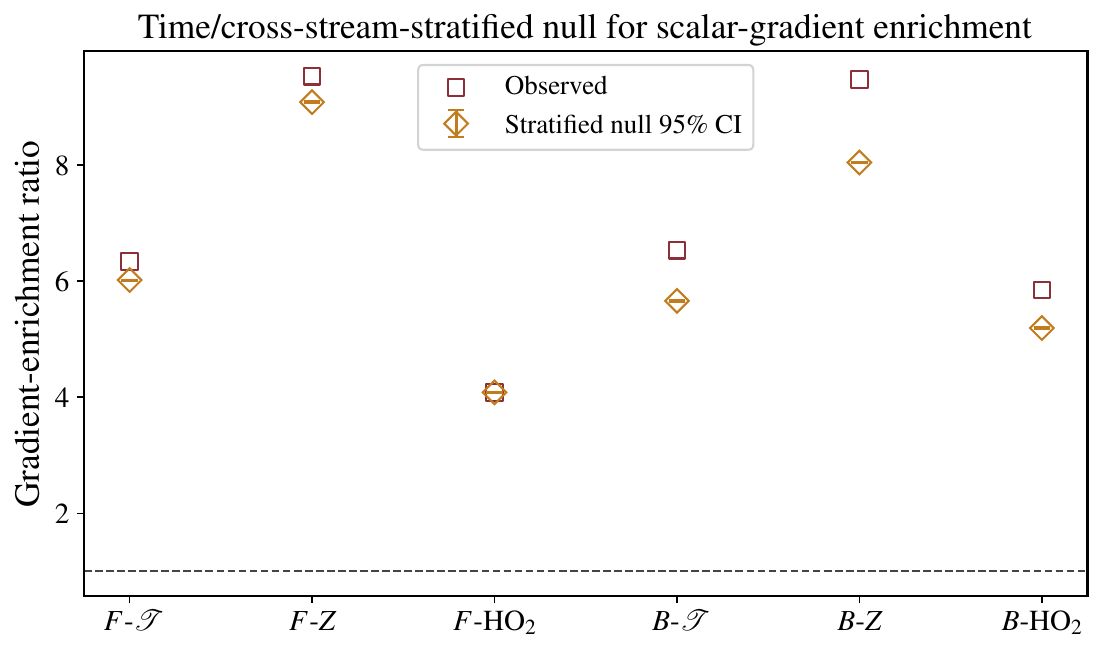}
  \caption{Time- and cross-stream-stratified null-model separation of ridge-conditioned scalar-gradient enrichment. Markers show observed ridge enrichment ratios; error bars show the stratified null mean and 95\% confidence interval.}
  \label{fig:ridge_scalar_null_significance}
\end{figure}

Figure~\ref{fig:ridge_scalar_null_significance} and Table~\ref{tab:ridge_scalar_null} show that the raw ridge enrichments remain large, with the stratified null explaining a substantial part of them. For forward ridges, the residual separation beyond the stratified null is modest for $\Temp$ and $Z$ ($R_{\mathrm{obs}}/R_{\mathrm{strat}}\approx1.05$) and essentially absent for HO$_2$ ($\approx1.00$). Backward ridges retain clearer residual separation, with $R_{\mathrm{obs}}/R_{\mathrm{strat}}\approx1.16$ for $\Temp$, $1.18$ for $Z$, and $1.13$ for HO$_2$. Ridge conditioning combines coherent-structure selectivity with shear-layer localization. The scalar conclusion is directional and conditional, with backward ridges retaining a measurable excess beyond time/cross-stream stratification and forward ridge enrichment dominated by the regions of the shear layer that the ridges occupy.

\input{tables/ridge_scalar_null_significance.tex}

\subsection{Lead/Lag Structure in Time-Resolved Ridge Activity}
To assess temporal ordering among coherent-structure activity measures, we compute detrended cross-correlations over the full overlap window using the forward ridge area fraction $A_f$, backward ridge area fraction $A_b$, forward/backward ridge-mean FTLE amplitudes ($S_f$, $S_b$), and forward/backward intersection area fraction $I_{fb}$. Cross-correlation is reported versus lag, with positive lag indicating that the first metric leads the second. Detrending removes numerical drift so that the estimated lags reflect fluctuation-scale ordering.

\begin{figure}[t]
  \centering
  \includegraphics[width=0.93\linewidth]{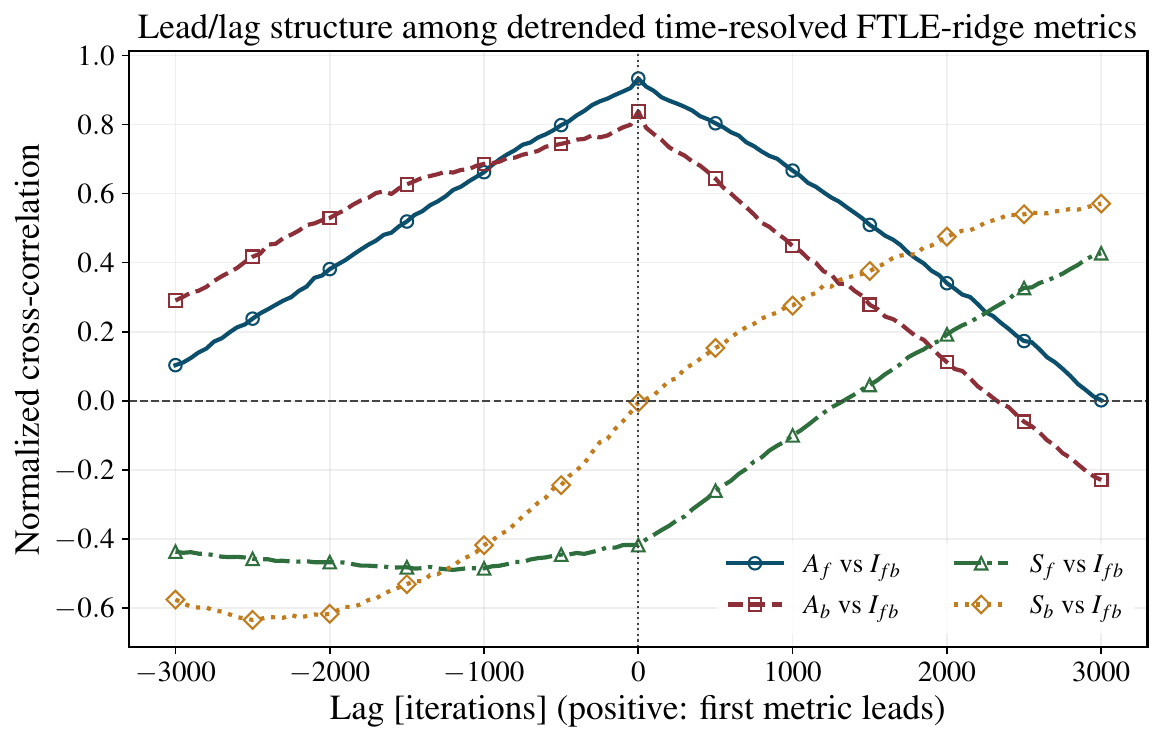}
  \caption{Detrended lead/lag cross-correlations among time-resolved FTLE-ridge activity metrics. Positive lag indicates the first metric leads, showing occupancy/strength phase relationships.}
  \label{fig:lcs_lead_lag}
\end{figure}

Figure~\ref{fig:lcs_lead_lag} and Table~\ref{tab:lcs_lead_lag} show that area-based activity is closely synchronized with ridge-intersection activity. $A_f$ vs $I_{fb}$ and $A_b$ vs $I_{fb}$ both peak at zero lag with strong positive correlation ($\approx0.93$ and $\approx0.84$, respectively). Strength-based metrics exhibit weaker, negative-lag anti-correlation at their peak absolute response ($S_f$ vs $I_{fb}$ and $S_b$ vs $I_{fb}$). With the stated sign convention, these negative lags mean that the intersection-area fluctuation leads the ridge-strength fluctuation for those peak anti-correlations. The implication is that overlap area tracks the geometric state nearly instantaneously, and mean ridge strength behaves as a phase-shifted response to that geometric state.

The dense scalar-coupling time series enable the same lead/lag framework for ridge-conditioned scalar metrics over the full overlap window, and the corresponding scalar-response analysis is assessed below in Sec.~\ref{sec:scalar_lead_lag}.

\input{tables/lead_lag_lcs_metrics.tex}

\subsection{Scalar-Response Lead/Lag Relative to Ridge Intersections}\label{sec:scalar_lead_lag}
We evaluate detrended lead/lag cross-correlations between intersection activity $I_{fb}$ and scalar enrichment ratios for $\Temp$, $Z$, and HO$_2$ in both forward and backward ridge families. Here, the enrichment is defined as
\begin{equation}
E_\phi = \frac{\langle |\nabla\phi|\rangle_{\mathrm{ridge}}}{\langle |\nabla\phi|\rangle_{\mathrm{all}}},
\end{equation}
and lag sign follows the convention that positive lag indicates $I_{fb}$ leads scalar response.

\begin{figure}[t]
  \centering
  \includegraphics[width=0.93\linewidth]{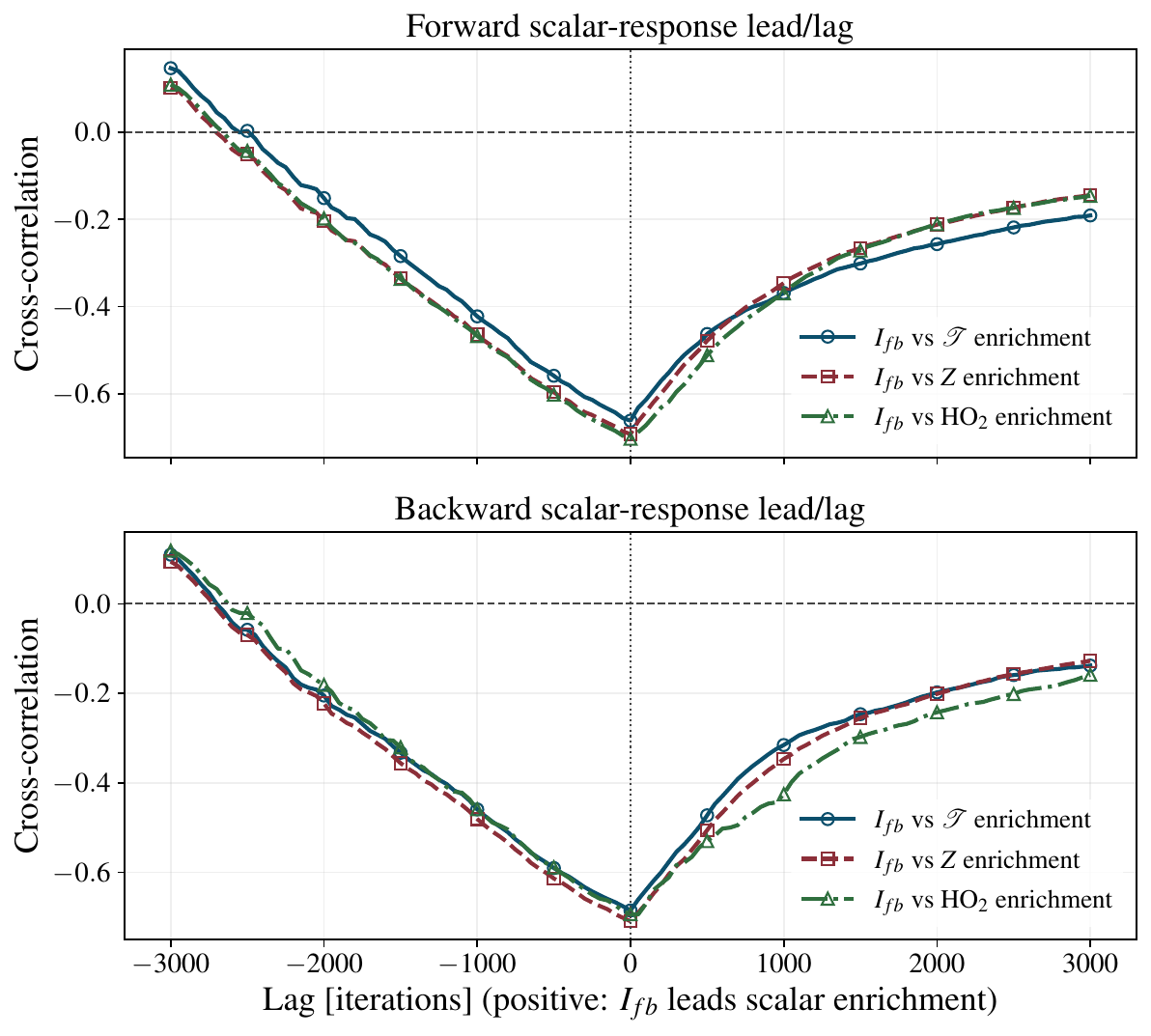}
  \caption{Detrended lead/lag correlations between intersection activity $I_{fb}$ and scalar enrichment ratios for $\Temp$, $Z$, and HO$_2$. Positive lag indicates $I_{fb}$ leads scalar response and tests temporal synchrony of transport coupling.}
  \label{fig:scalar_lead_lag}
\end{figure}

Figure~\ref{fig:scalar_lead_lag} and Table~\ref{tab:scalar_lead_lag} show strong peak absolute correlations that are predominantly near zero lag for all three scalars. For forward ridges, peak values are approximately $-0.66$ ($\Temp$), $-0.69$ ($Z$), and $-0.70$ (HO$_2$), all at zero lag. For backward ridges, $\Temp$ and $Z$ similarly peak at zero lag (approximately $-0.69$ and $-0.71$). HO$_2$ shows a marginal one-step offset (50 iterations). The near-zero-lag structure indicates quasi-synchronous evolution of ridge-intersection fluctuations and scalar-enrichment fluctuations on resolved timescales. The negative sign reflects anti-phase detrended behavior, with intervals of elevated overlap activity coinciding with local decreases in enrichment relative to each series' trend.

\input{tables/lead_lag_scalar_response.tex}

\subsection{Robustness}
This subsection assesses the consistency of the directional ridge asymmetry and lag-based transport inferences under geometric, topological, and timescale changes. These analyses directly constrain the main transport conclusions. The resampling-heavy and region-conditioned extensions are given in Appendix~\ref{app:supp_robustness}.

\begin{figure}[t]
  \centering
  \includegraphics[width=0.96\linewidth]{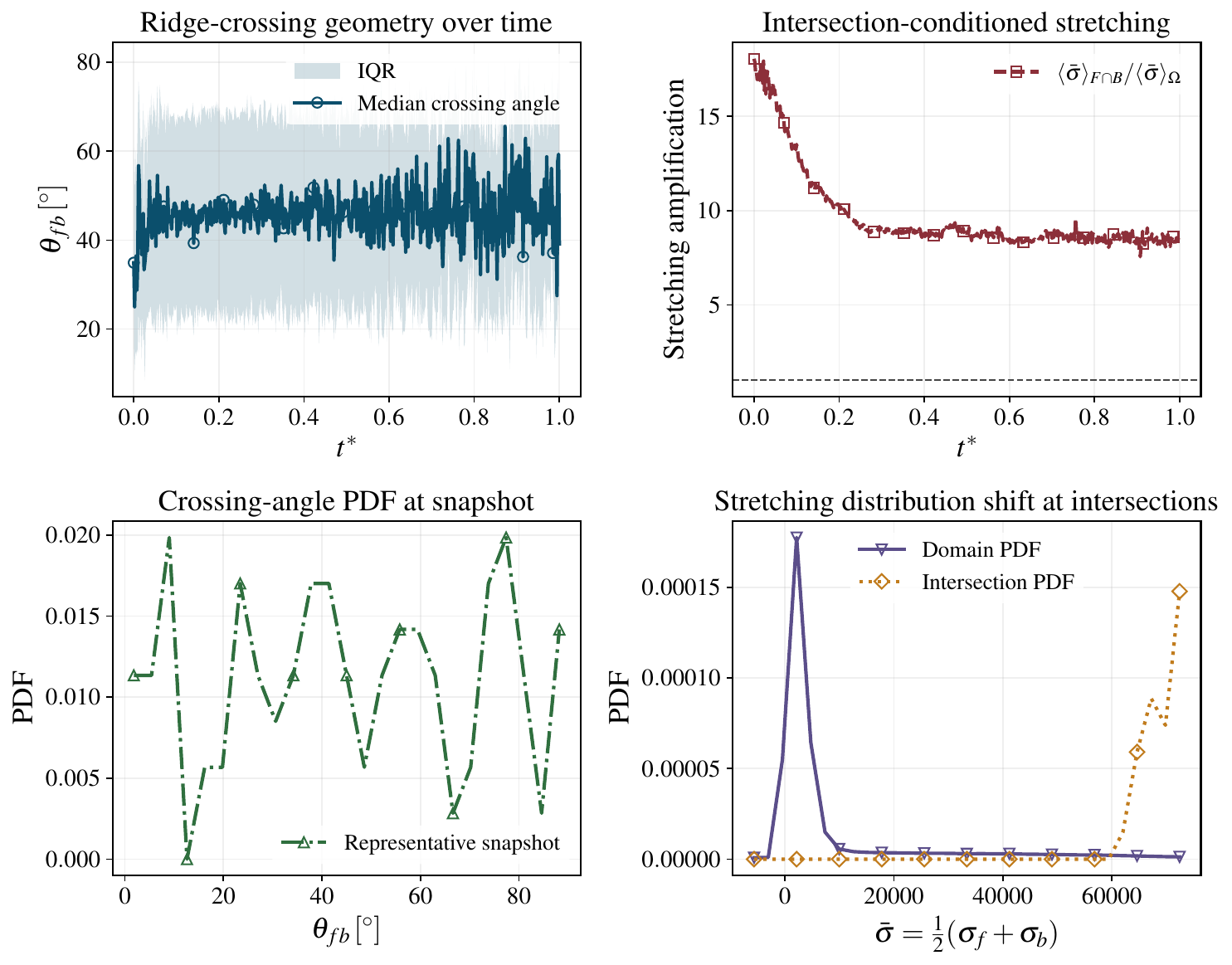}
  \caption{FTLE-ridge geometry from time-resolved ridge-crossing angle statistics, intersection-conditioned stretching amplification, and instantaneous PDFs at a representative time.}
  \label{fig:lcs_geometry_diag}
\end{figure}

\paragraph{FTLE-ridge geometry.}
To provide an assessment beyond aggregate FTLE-ridge envelope behavior, we quantify forward/backward ridge-crossing geometry and intersection-conditioned stretching (Fig.~\ref{fig:lcs_geometry_diag}, Table~\ref{tab:lcs_geometry_diag}). The crossing-angle measure is
\begin{align}
\theta_{fb}=\cos^{-1}\!\left(\left|\hat{\mathbf{n}}_f\!\cdot\!\hat{\mathbf{n}}_b\right|\right)
\end{align}
tracks ridge-crossing transversality over time. The amplification ratio
\begin{align}
    A_s = \langle\bar{\sigma}\rangle_{F\cap B}/\langle\bar{\sigma}\rangle_{\Omega};\quad \bar{\sigma}=\tfrac{1}{2}(\sigma_f+\sigma_b)
\end{align}
measures how strongly finite-time stretching concentrates at intersection sets. Together, these quantities distinguish simple overlap from dynamically significant crossing events. The reported values show broad, consistently oblique crossing geometry. The time-averaged median crossing angle is $46.13^\circ$, with a mean interquartile width of $44.22^\circ$ and an instantaneous median of $43.69^\circ$. The stretching amplification is also large, with $\langle A_s\rangle\approx9.70$ and an instantaneous representative value of $8.94$. Intersection sets coincide with regions where the local finite-time stretching is roughly an order of magnitude stronger than the domain average, marking dynamically active planar transport regions.

\input{tables/lcs_geometry_diagnostics.tex}

\begin{figure}[t]
  \centering
  \includegraphics[width=0.98\linewidth]{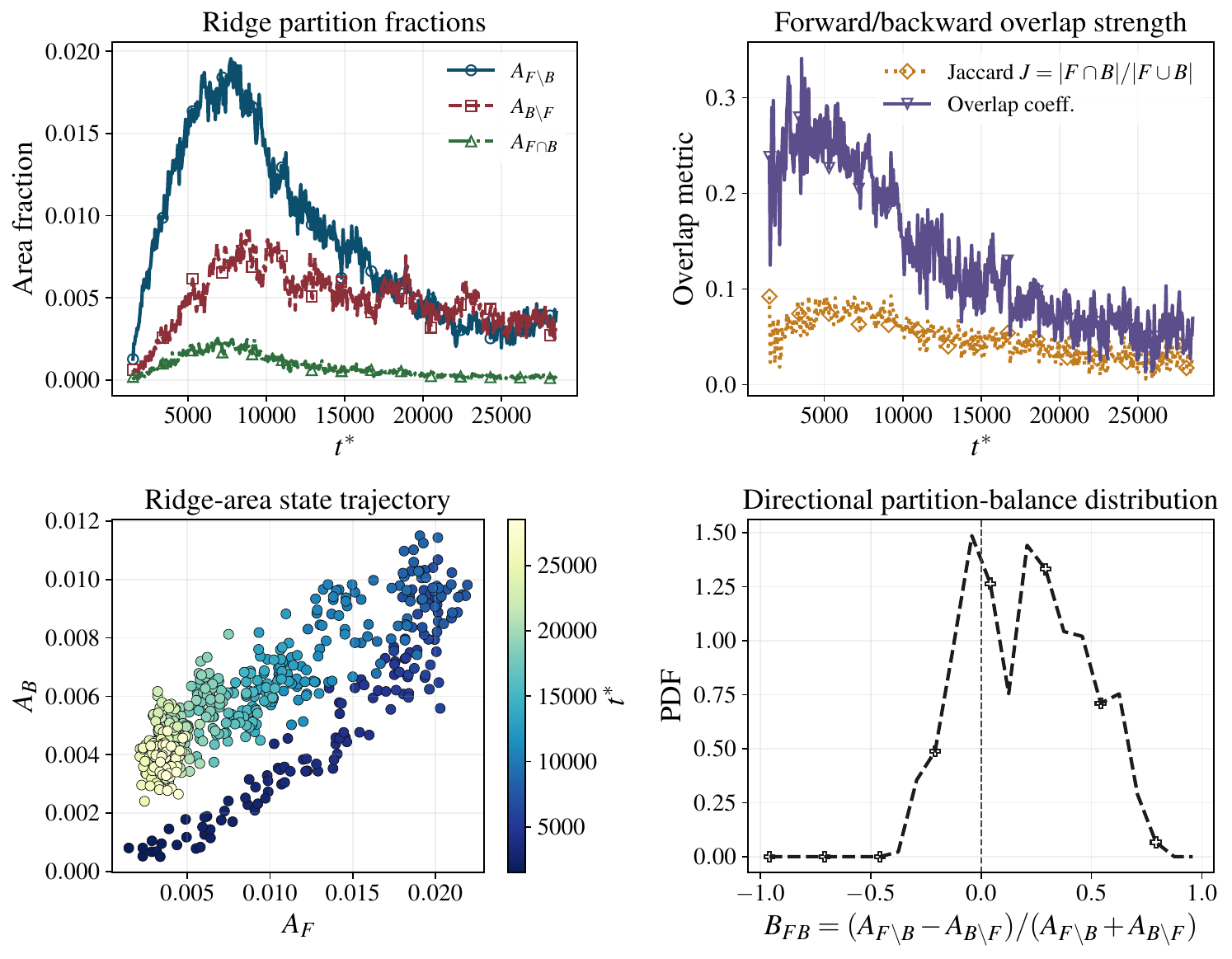}
  \caption{FTLE-ridge topology from partition-area fractions, overlap indices, area-state trajectory, and partition-balance distribution.}
  \label{fig:lcs_topology_diag}
\end{figure}

\paragraph{Topology.}
We partition ridge support into direction-specific and shared regions over time (Fig.~\ref{fig:lcs_topology_diag}, Table~\ref{tab:lcs_topology_diag}) using forward-only ($F\setminus B$), backward-only ($B\setminus F$), and intersection ($F\cap B$) regions. The partition measures how much of the extracted skeleton is direction-specific versus mutually supported by attracting/repelling ridge families. Overlap indices, measured as the Jaccard overlap $J=|F\cap B|/|F\cup B|$ and overlap coefficient, track bidirectional co-organization strength. The partition-balance index $B_{FB}$ further exposes persistent directional bias in ridge occupancy. The tabulated areas show that the topology is dominated by direction-specific support. The means are $\langle A_{F\setminus B}\rangle\approx8.50\times10^{-3}$ and $\langle A_{B\setminus F}\rangle\approx4.89\times10^{-3}$, with mean intersection area $\langle A_{F\cap B}\rangle\approx7.88\times10^{-4}$. Consistent with that separation, the mean Jaccard overlap is $0.0476$ and its maximum is $0.0932$, so forward and backward ridge families occupy predominantly distinct sets. The positive mean balance index, $\langle B_{FB}\rangle\approx0.200$, indicates a persistent forward-area bias, and its standard deviation of $0.262$ shows that the bias varies in strength. The topology is consistent with the earlier geometric asymmetry. The two ridge families are co-organized enough to generate localized intersection hotspots and remain predominantly distinct with a clear directional imbalance.

\input{tables/lcs_topology_diagnostics.tex}

\begin{figure}[t]
  \centering
  \includegraphics[width=0.95\linewidth]{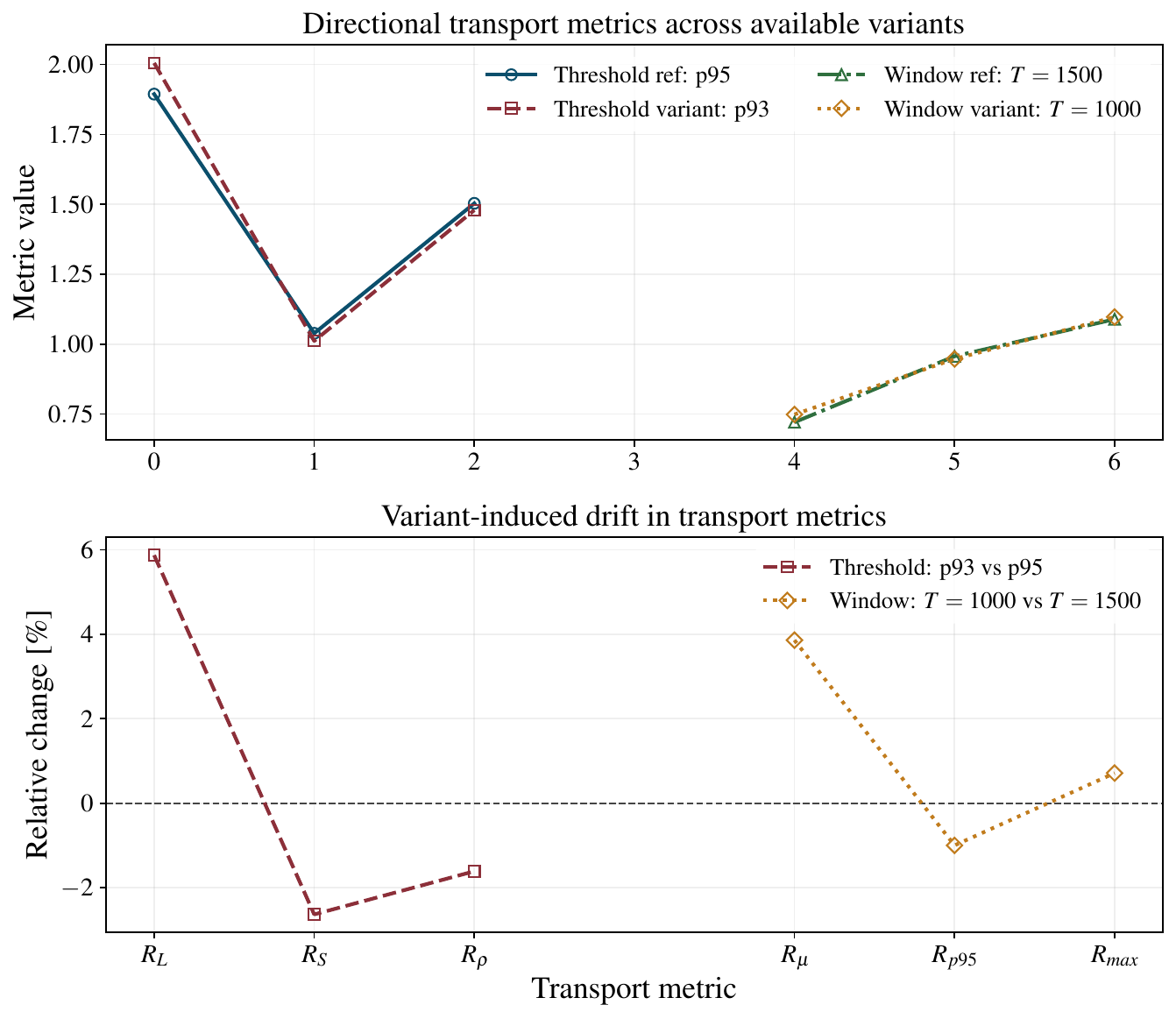}
  \caption{Sensitivity across threshold and integration-window variants. Upper panels show absolute metric values by variant, and lower panels show relative drift from each group's reference variant.}
  \label{fig:conclusion_metric_sensitivity}
\end{figure}

\begin{figure}[t]
  \centering
  \includegraphics[width=0.95\linewidth]{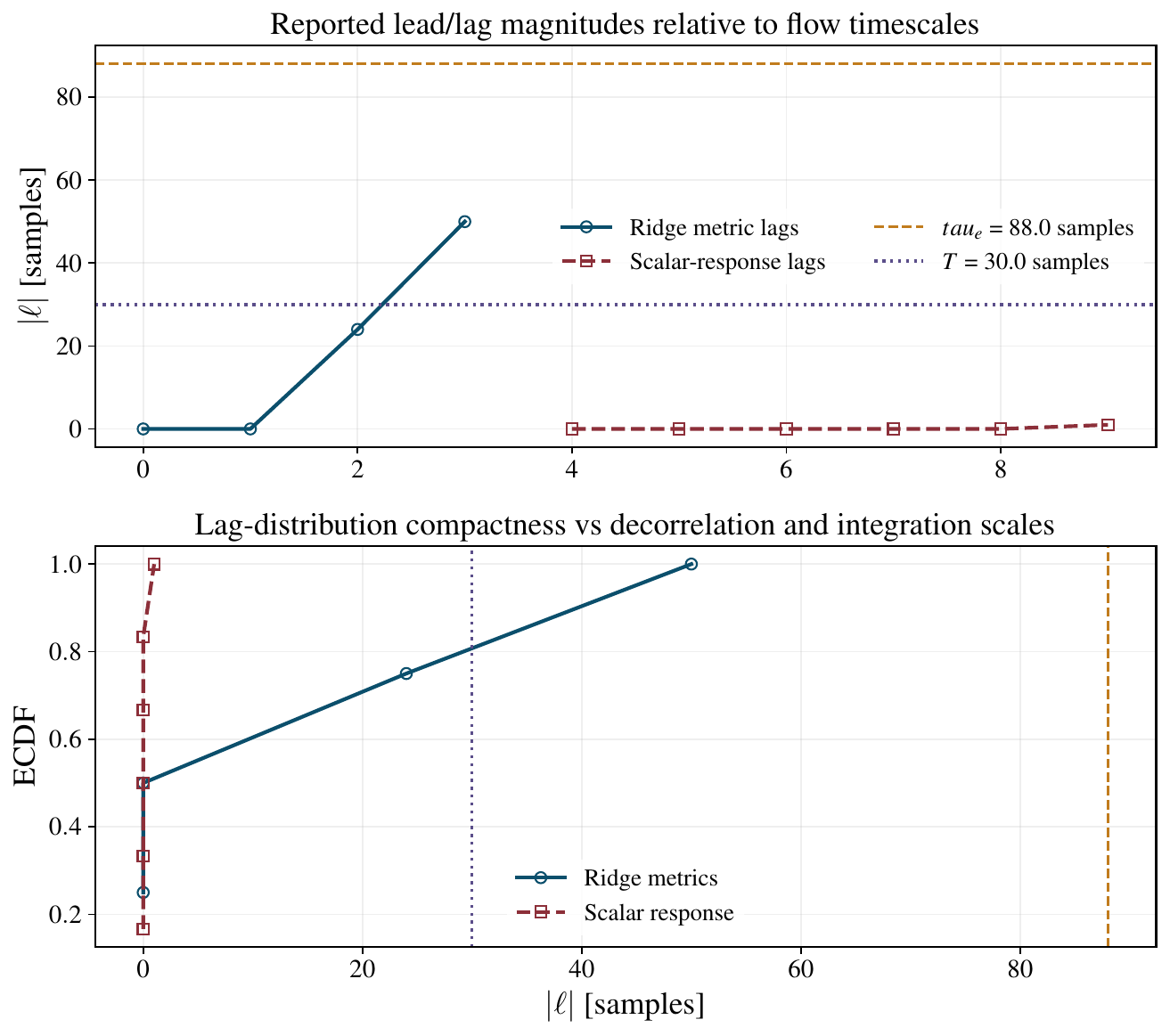}
  \caption{Timescale consistency for reported lags. Upper panels show reported lag magnitudes with decorrelation/integration reference levels, and lower panels show lag ECDFs for ridge and scalar-response metrics.}
  \label{fig:timescale_consistency}
\end{figure}

\paragraph{Sensitivity and timescale closure.}
Fig.~\ref{fig:conclusion_metric_sensitivity} and Table~\ref{tab:conclusion_metric_sensitivity} report compact variant analyses for directional ridge-asymmetry and directional FTLE ratios under the available threshold and integration-window alternatives. Under threshold variation, the key ratios that influence the major observations move little. The length-asymmetry ratio increases from $1.894$ to $2.005$ ($+5.9\%$), the strength ratio shifts from $1.039$ to $1.011$ ($-2.6\%$), and the peak column-occupancy ratio changes from $1.503$ to $1.479$ ($-1.6\%$). Under window variation, the forward/backward FTLE ratios are similarly stable; the mean-strength ratio changes by $3.9\%$, the $p95$ ratio by $-1.0\%$, and the max-ratio by $0.7\%$. These analyses indicate that the directional conclusions are insensitive to the available threshold and integration-window alternatives within the tested range. Figure~\ref{fig:timescale_consistency} and Table~\ref{tab:timescale_consistency} then place lead/lag magnitudes against decorrelation and integration scales, so lag magnitudes are assessed within an explicit temporal hierarchy. The scalar-response lags are especially compact, with median absolute lag equal to zero samples. The ridge-metric lags have a median of 12 samples. Both remain small relative to the FTLE integration window of 30 samples and far below the $I_{fb}$ decorrelation scale of 88 samples in median terms. All reported lags satisfy $|\ell|\le\tau_e$, and $90\%$ satisfy $|\ell|\le T$, supporting the reported synchrony and offset behavior as resolved temporal associations at the analysis-window scale. These analyses bound the temporal consistency of the correlations and leave mechanistic sequencing as a separate question.

\input{tables/conclusion_metric_sensitivity.tex}

\input{tables/timescale_consistency.tex}

\subsection{Uncertainty Envelopes Across Variants}
\begin{figure}[t]
  \centering
  \includegraphics[width=0.95\linewidth]{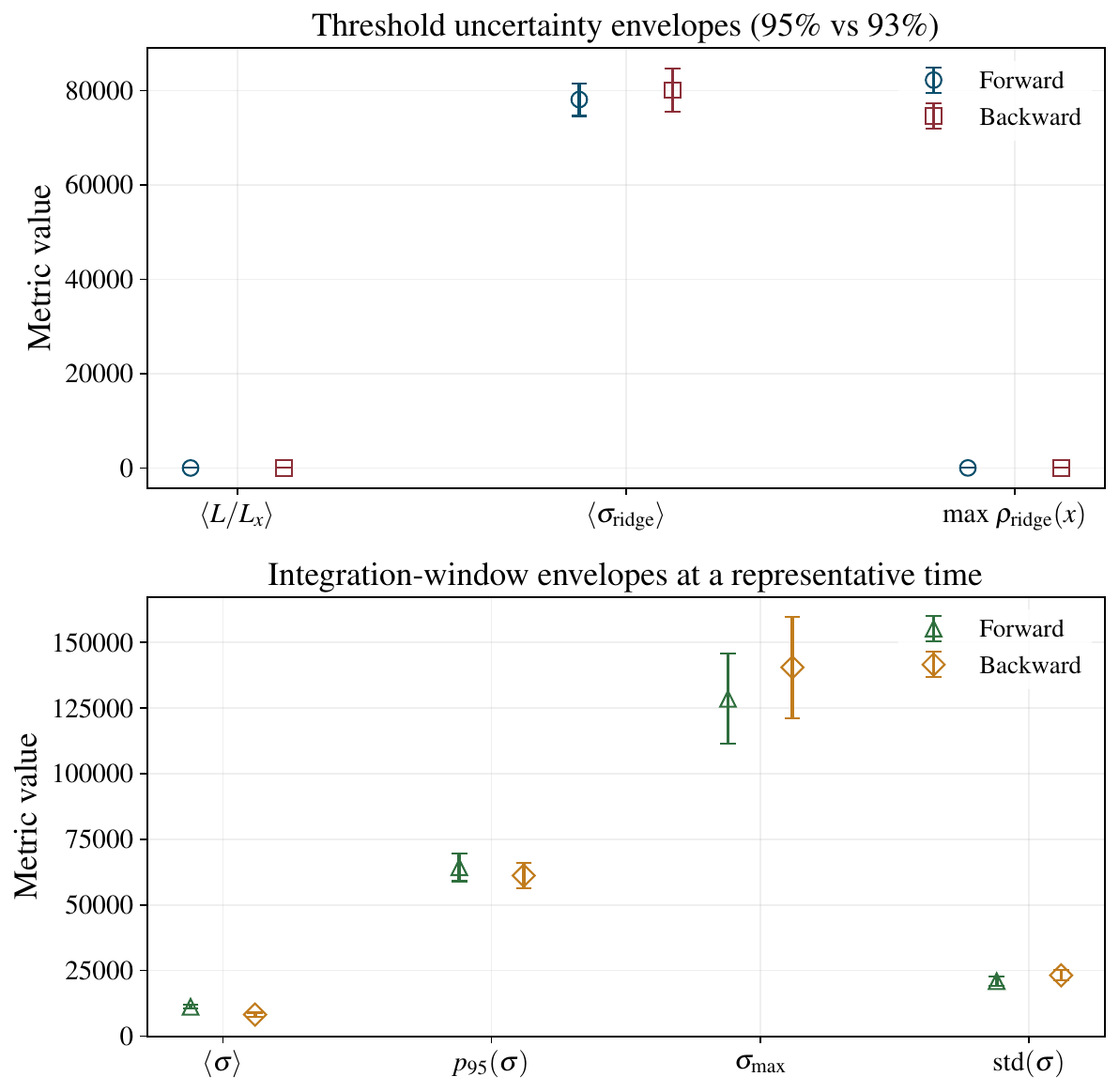}
  \caption{Uncertainty envelopes for key ridge and FTLE metrics across available parameter variants. Upper panels show threshold variation (95\% vs 93\%), and lower panels show integration-window variation at a matched representative time.}
  \label{fig:uncertainty_envelopes}
\end{figure}

To quantify parameter sensitivity in a compact way, Fig.~\ref{fig:uncertainty_envelopes} and Table~\ref{tab:uncertainty_envelope} report min--max envelopes across ridge extraction and integration variants. For ridge extraction, we use percentile thresholds of 95\% and 93\% for both directions. For integration-window sensitivity, we compare the shorter and longer windows at the same representative time within their overlapping availability. This keeps the comparison data-consistent and provides a quantitative spread estimate.

The threshold envelopes show that directional ordering is stable under cutoff variation; the forward ridges remain longer on average than backward ridges, and backward ridges retain comparable-to-higher mean ridge strength. Relative spreads are moderate for $\langle L/L_x\rangle$ and $\langle\sigma_{\mathrm{ridge}}\rangle$ (roughly $9$--$24$\%), with larger spreads for peak streamwise column occupancy (roughly $42$--$43$\%), indicating that local occupancy peaks are more threshold-sensitive than bulk ridge-length or ridge-strength means. For window sensitivity at the representative time, forward/backward FTLE means and $p_{95}$ levels vary by approximately $14$--$18$\%. Peak and standard-deviation metrics vary by approximately $17$--$27$\%. These spreads warrant explicit reporting but preserve the directional asymmetry and scalar-coupling conclusions in this work.

\input{tables/uncertainty_envelope_summary.tex}

\subsection{Planar scope}
The analysis is intentionally planar. Particle trajectories are advanced with the in-plane velocity components on the mid-$z$ slice, so the extracted FTLE ridges describe finite-time transport in that constrained slice. Connecting these curves to material surfaces in the full three-dimensional DNS would require a separate three-dimensional extraction. The out-of-plane velocity comparison in Table~\ref{tab:out_of_plane_velocity} shows that $w$ is small relative to in-plane RMS velocity over the full plane (median ratio $\approx0.04$) and larger on the ridge union (median ratio $\approx0.19$). The mid-plane calculation captures a meaningful in-plane transport skeleton and samples three-dimensional motion near the most active ridge regions.

The main statistical backbone is an operational FTLE-ridge skeleton obtained from thresholding, skeletonization, and component filtering. These masks are available over the synchronized time sequence required for ridge geometry, intersections, scalar conditioning, null tests, and lead/lag analysis. The hyperbolic geodesic-LCS statistics are computed from Cauchy--Green tensors reconstructed from the same planar velocity sequence at a coarser start-time stride, giving 55 start times per family across the synchronized interval. The paper combines dense FTLE-ridge statistics with a time-resolved variational/geodesic material-curve sample at a coarser cadence. Extending the geodesic-LCS extraction to every 50-iteration synchronized ridge time would require reconstructing Cauchy--Green tensor fields, or equivalently the underlying flow maps, at those same starts. The threshold and integration-window envelopes evaluate the reported directional trends under the available variants. Finally, the stratified null model controls time and cross-stream shear-layer localization, which is stricter than an area-only random mask. Other possible covariates of scalar-gradient intensity remain outside the null construction. The scalar-coupling results provide evidence for FTLE-ridge-conditioned planar organization beyond the tested nulls, with chemistry represented through conditional scalar-gradient response in the analyzed plane.

\section{Conclusions}
The results support three main conclusions about planar Lagrangian transport in the reacting temporal mixing layer. First, the mid-plane forward/backward transport skeleton is directionally asymmetric and spatially localized. Forward ridges are longer and more streamwise concentrated, backward ridges carry stronger spectral amplitude, and forward/backward intersections occupy small recurrent hotspots with oblique crossing geometry and roughly order-of-magnitude stretching amplification. The planar geodesic-LCS extraction supports this result. Selected strainlines occupy the same high-strain scaffold and remain more selective than the dense operational FTLE-ridge masks, and the one-factor geodesic sensitivity analysis shows persistence across seed-percentile, retained-curve-cap, and support-threshold variations.

Second, ridge conditioning acts as a strong scalar-organization filter. Ridge-conditioned gradient enrichment is large for $\Temp$, $Z$, and HO$_2$, with the strongest amplification in mixture-fraction gradients. The time- and cross-stream-stratified null model shows that much of the raw enrichment is inherited from shear-layer localization. The scalar result is directional and conditional, with backward ridges retaining clearer residual scalar-gradient excess beyond the tested null and forward-ridge enrichment largely explained by where those ridges sit in the shear layer. These results support coherent-structure-mediated scalar organization in the planar analysis, with chemistry assessed through conditional scalar-gradient response in the mid-plane.

Third, the temporal statistics indicate compact fluctuation-level coupling with no stationary lead/lag hierarchy. Ridge-area fluctuations are nearly synchronous with intersection activity, and scalar-enrichment fluctuations show strong near-zero-lag anti-phase correlations with ridge overlap. The lag-timescale checks place these lags below the relevant decorrelation and FTLE-integration scales, supporting the measured synchrony at the analysis-window scale. Time-segment measures show appreciable nonstationarity in overlap and scalar-enrichment levels. The resulting picture is finite-time and planar. Coherent structures repeatedly organize where scalar gradients concentrate and redistribute, with relationships conditioned on the mid-plane surrogate and the analyzed time interval.

\begin{acknowledgments}
The computing power for this study was provided by the Phoenix Computing Cluster as part of Georgia Tech's Partnership for Advanced Computing Environment, and is gratefully acknowledged.
\end{acknowledgments}

\section*{Author Declarations}

\subsection*{Author Contributions}
Sriram P. Kalathoor: Conceptualization, Methodology, Software, Formal analysis, Investigation, Visualization, Writing -- original draft. Joseph C. Oefelein: Conceptualization, Resources, Supervision, Writing -- review and editing.

\subsection*{Funding}
This research received no specific grant from any funding agency, commercial or not-for-profit sectors.

\subsection*{Conflict of Interest}
The authors have no conflicts to disclose.

\section*{Data Availability}
The data that support the findings of this study are available from the corresponding author upon reasonable request.

\appendix
\section{Supplementary Robustness Measures}\label{app:supp_robustness}
This appendix presents supporting robustness measures that complement the primary conclusions in the main part of this work. These include event-conditioned composites, bootstrap lag intervals, region-conditioned scalar budgets, asymmetry indices, projection-based null tests, and time-segment robustness checks. Their role is to quantify second-order uncertainty and conditioning effects after the main FTLE-ridge geometry/topology and geodesic-LCS analyses have established directional asymmetry and transport coupling. The appendix is organized from conditioning/lag uncertainty to region-partition behavior and statistical/temporal robustness, following the same sequence of structure, response characterization, and robustness closure used in the main text.

\paragraph{Event-conditioned composites.}
Using top-10\% and bottom-10\% intersection-activity windows, Table~\ref{tab:event_composites} reports how scalar enrichment and absolute ridge-normal gradient projection respond during high-overlap episodes. The response is scalar- and direction-dependent, indicating a non-universal scalar-response mode during overlap bursts. Quantitatively, forward-$\Temp$ enrichment increases mildly ($E_{high}/E_{low}\approx1.07$). Both forward-$Z$ and backward-$Z$ decrease ($\approx0.96$ and $\approx0.82$), and HO$_2$ decreases in both directions ($\approx0.92$ forward, $\approx0.90$ backward).

\paragraph{Bootstrap lead/lag uncertainty.}
The detrended lead/lag peak locations are accompanied by broad block-bootstrap confidence intervals (Table~\ref{tab:leadlag_bootstrap}), so point-estimated lag extrema are weakly localized in lag even when peak-correlation magnitudes are substantial. In particular, all reported lag confidence intervals span nearly the full tested range (about $[-60,60]$ steps). Peak correlations remain consistently moderate/strong (about $-0.66$ to $-0.71$), supporting strong coupling magnitude and weakly constrained exact lag location.

\paragraph{Budgets.}
Table~\ref{tab:intersection_budgets} gives scalar-gradient means for the $F_{\mathrm{only}}$, $B_{\mathrm{only}}$, and $F\cap B$ partitions. Table~\ref{tab:asymmetry_indices} compresses directional differences into asymmetry indices $A=(M_f-M_b)/(M_f+M_b)$. Together, these measures separate spatial partition effects from directional aggregate effects. For $\Temp$ and $Z$, intersection values exceed both one-sided regions ($\langle|\nabla \Temp|\rangle_{F\cap B}\approx2.73\times10^{5}$ and $\langle|\nabla Z|\rangle_{F\cap B}\approx1.62\times10^{2}$), consistent with stronger gradient localization where forward and backward ridge families co-occur. HO$_2$ is different in that $B_{\mathrm{only}}$ is largest ($\approx1.03\times10^{-1}$), indicating species-dependent conditioning. The asymmetry is consistent with the geometric observations. The length and ridge-area indices are positive ($A\approx0.309$ and $0.241$), and strength and enrichment indices are mildly negative.

\paragraph{Projection null tests.}
In addition to gradient-enrichment null testing, Table~\ref{tab:projection_stratified_null} and Fig.~\ref{fig:app_supp_b}(a) apply the same time- and cross-stream-stratified null to ridge-conditioned absolute scalar-gradient projection ratios. Time-segment robustness measures (Table~\ref{tab:segment_robustness}, Fig.~\ref{fig:app_supp_b}(b)) quantify how key metrics evolve across early/middle/late windows, separating persistent directional tendencies from metrics with stronger nonstationary drift. The projection results mirror the scalar-gradient enrichment results. Raw ridge-conditioned projection ratios are large ($R_{obs}\sim4.1$--$9.6$), and much of that amplification is expected from the ridge families' time-varying cross-stream support. Residual projection separation is strongest for backward ridges ($R_{obs}/R_{strat}\approx1.13$--$1.18$), modest for forward $\Temp$ and $Z$ ($\approx1.05$), and nearly absent for forward HO$_2$. Segment measures show notable nonstationarity in overlap/area and scalar-enrichment metrics, with $I_{fb}$ and $E_Z$ spreads near or above 100\%. Ridge-strength metrics vary much less (about 15--18\%), indicating that structural intensity is more stable than overlap-driven occupancy/enrichment levels.

\begin{table}[t]
  \centering
  \caption{Event-conditioned composite means (bottom 10\% vs top 10\% of intersection activity).}
  \label{tab:event_composites}
  \scriptsize
  \setlength{\tabcolsep}{2.4pt}
  \begin{tabular}{llllll}
    \hline
    Direction & Scalar & $E_{low}$ & $E_{high}$ & $E_{high}/E_{low}$ & $G_{high}/G_{low}$ \\
    \hline
    forward & $\Temp$ & 6.507 & 6.961 & 1.070 & 1.490 \\
    forward & $Z$ & 10.577 & 10.154 & 0.960 & 1.305 \\
    forward & $\mathrm{HO_2}$ & 10.556 & 9.712 & 0.920 & 0.181 \\
    backward & $\Temp$ & 8.486 & 7.693 & 0.907 & 1.334 \\
    backward & $Z$ & 13.191 & 10.797 & 0.819 & 1.158 \\
    backward & $\mathrm{HO_2}$ & 10.981 & 9.869 & 0.899 & 0.165 \\
    \hline
  \end{tabular}
\end{table}

\begin{table}[t]
  \centering
  \caption{Block-bootstrap confidence intervals for scalar lead/lag peak correlations.}
  \label{tab:leadlag_bootstrap}
  \tiny
  \setlength{\tabcolsep}{1.6pt}
  \begin{tabular}{llllll}
    \hline
    Direction & Scalar & Peak corr. & 95\% CI (corr) & Peak lag [steps] & 95\% CI (lag) \\
    \hline
    forward & $\Temp$ & -0.662 & [-0.412,0.813] & 0 & [-60.0,60.0] \\
    forward & $Z$ & -0.693 & [-0.404,0.780] & 0 & [-60.0,60.0] \\
    forward & $\mathrm{HO_2}$ & -0.703 & [-0.385,0.780] & 0 & [-60.0,60.0] \\
    backward & $\Temp$ & -0.685 & [-0.367,0.748] & 0 & [-60.0,60.0] \\
    backward & $Z$ & -0.709 & [-0.398,0.797] & 0 & [-60.0,60.0] \\
    backward & $\mathrm{HO_2}$ & -0.694 & [-0.381,0.714] & 1 & [-60.0,60.0] \\
    \hline
  \end{tabular}
\end{table}

\begin{table}[!htbp]
  \centering
  \caption{Region-conditioned scalar-gradient means for forward-only, backward-only, and intersection regions.}
  \label{tab:intersection_budgets}
  \scriptsize
  \begin{tabular}{llll}
    \hline
    Region & Scalar & $\langle|\nabla\phi|\rangle$ & Samples \\
    \hline
    $F_{\mathrm{only}}$ & $\Temp$ & $2.22\times10^{5}$ & 1204934 \\
    $B_{\mathrm{only}}$ & $\Temp$ & $2.28\times10^{5}$ & 692941 \\
    $F\cap B$ & $\Temp$ & $2.73\times10^{5}$ & 111769 \\
    $F_{\mathrm{only}}$ & $Z$ & $1.33\times10^{2}$ & 1204934 \\
    $B_{\mathrm{only}}$ & $Z$ & $1.31\times10^{2}$ & 692941 \\
    $F\cap B$ & $Z$ & $1.62\times10^{2}$ & 111769 \\
    $F_{\mathrm{only}}$ & $\mathrm{HO_2}$ & $6.87\times10^{-2}$ & 1204934 \\
    $B_{\mathrm{only}}$ & $\mathrm{HO_2}$ & $1.03\times10^{-1}$ & 692941 \\
    $F\cap B$ & $\mathrm{HO_2}$ & $6.63\times10^{-2}$ & 111769 \\
    \hline
  \end{tabular}
\end{table}

\begin{table}[!htbp]
  \centering
  \caption{Forward/backward asymmetry indices $A=(M_f-M_b)/(M_f+M_b)$ for key metrics.}
  \label{tab:asymmetry_indices}
  \scriptsize
  \begin{tabular}{llll}
    \hline
    Metric & $M_f$ & $M_b$ & $A$ \\
    \hline
    $\langle L/L_x\rangle$ & 0.1779 & 0.09394 & 0.309 \\
    $A_{\mathrm{ridge}}$ & 0.009284 & 0.005674 & 0.241 \\
    $\langle\sigma_{\mathrm{ridge}}\rangle$ & 7.944e+04 & 8.46e+04 & -0.032 \\
    $E_{\Temp}$ & 6.527 & 7.97 & -0.100 \\
    $E_Z$ & 10.31 & 12.2 & -0.084 \\
    $E_{\mathrm{HO_2}}$ & 10.19 & 10.79 & -0.028 \\
    \hline
  \end{tabular}
\end{table}

\input{tables/ridge_projection_stratified_null.tex}

\begin{center}
  \captionsetup{type=table}
  \centering
  \captionof{table}{Time-segment robustness of key metrics (early/middle/late thirds of the common time window).}
  \label{tab:segment_robustness}
  \scriptsize
  \begin{tabular}{lllll}
    \hline
    Metric & Early & Middle & Late & Spread [\%] \\
    \hline
    $I_{fb}$ & 0.001458 & 0.000693 & 0.0002163 & 148.3 \\
    $A_f$ & 0.0152 & 0.009003 & 0.003681 & 122.0 \\
    $A_b$ & 0.006555 & 0.006263 & 0.004212 & 43.5 \\
    $S_f$ & 8.701e+04 & 7.679e+04 & 7.454e+04 & 15.4 \\
    $S_b$ & 9.389e+04 & 8.165e+04 & 7.831e+04 & 18.1 \\
    $E_{f,Z}$ & 17.87 & 7.071 & 6.02 & 99.2 \\
    $E_{b,Z}$ & 22.14 & 7.664 & 6.814 & 105.9 \\
    \hline
  \end{tabular}
\end{center}

\begin{center}
  \begin{minipage}{\linewidth}
  \captionsetup{type=figure}
  \centering
  \begin{subfigure}{0.9\linewidth}
    \centering
    \includegraphics[width=\linewidth]{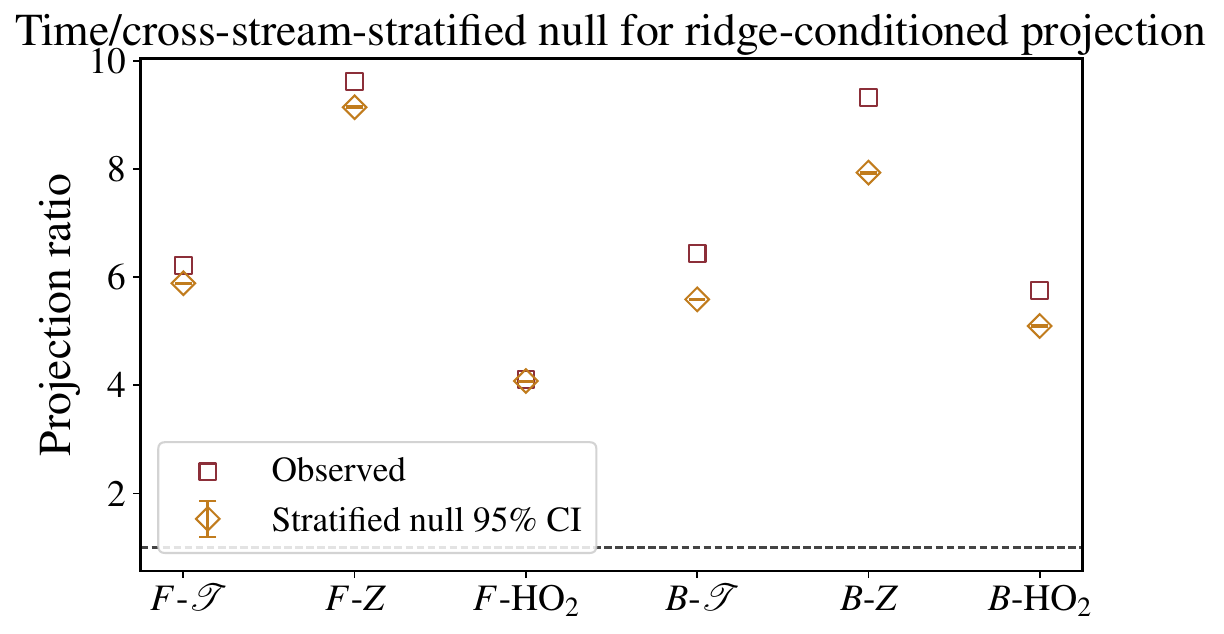}
    \caption{Projection-ratio null separation.}
  \end{subfigure}
  \par\smallskip
  \begin{subfigure}{0.8\linewidth}
    \centering
    \includegraphics[width=\linewidth]{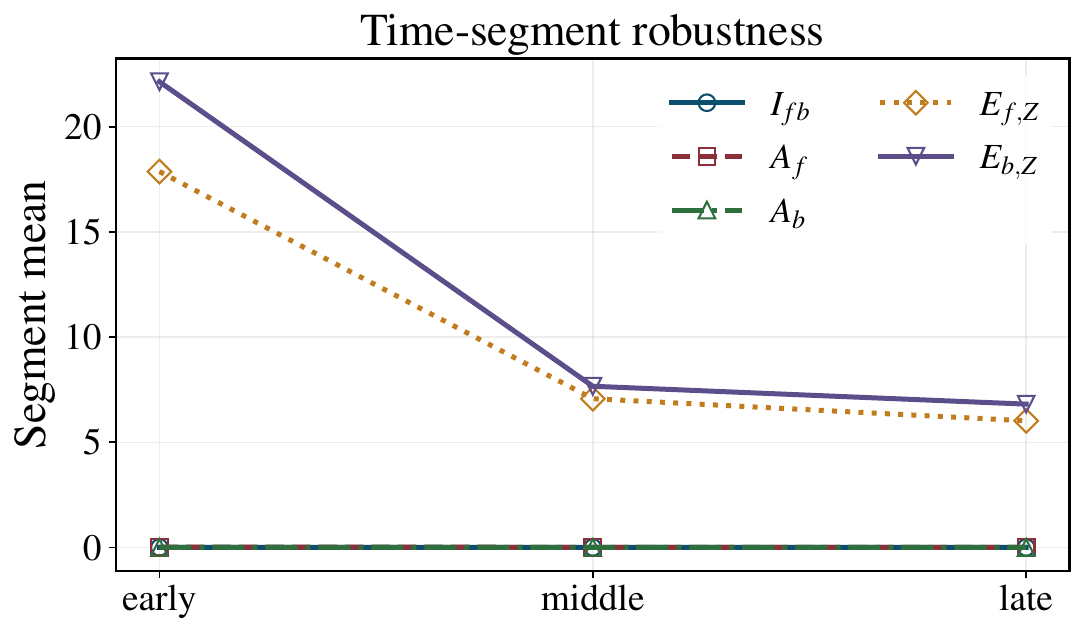}
    \caption{Time-segment robustness.}
  \end{subfigure}
  \addtocounter{figure}{-1}
  \captionof{figure}{Supplementary robustness assessment showing null-separation and segment-robustness views.}
  \label{fig:app_supp_b}
  \end{minipage}
\end{center}

\bibliographystyle{aipauth4-1}
\bibliography{refs}

\end{document}

%% file: tables/out_of_plane_velocity_summary.tex
\begin{table}[t]
  \centering
  \caption{Out-of-plane velocity magnitude relative to in-plane speed on the analyzed mid-plane slice. Ratios are RMS values over time.}
  \label{tab:out_of_plane_velocity}
  \begin{tabular}{llll}
    \hline
    Region & Mean & Median & 10--90\% range \\
    \hline
    Full plane & 0.040 & 0.040 & [0.034,0.048] \\
    $F\cup B$ ridges & 0.196 & 0.191 & [0.159,0.247] \\
    \hline
  \end{tabular}
\end{table}

%% file: tables/ridge_stats_summary.tex
\begin{table*}[t]
\centering
\caption{Ridge statistics summary. Lengths are nondimensionalized by $L_x$ and estimated from skeleton arclength using the local grid spacing.}
\label{tab:ridge_stats}
\begin{tabular}{llcccc}
\hline
Direction & Metric & Mean & Median & P10 & P90 \\
\hline
Forward & Ridge length $L/L_x$ & 0.178 & 0.0684 & 0.041 & 0.348 \\
Forward & Ridge strength (FTLE) & 8.16e+04 & 7.93e+04 & 6.43e+04 & 1.02e+05 \\
Backward & Ridge length $L/L_x$ & 0.0939 & 0.0625 & 0.041 & 0.18 \\
Backward & Ridge strength (FTLE) & 8.47e+04 & 8.23e+04 & 6.25e+04 & 1.1e+05 \\
\hline
\end{tabular}
\end{table*}

%% file: tables/geodesic_lcs_timeseries_summary.tex
\begin{table*}[t]
  \centering
  \caption{Time-resolved planar hyperbolic geodesic-LCS statistics computed from Cauchy--Green tensors reconstructed over the available planar velocity sequence. Curve lengths are nondimensionalized by $L_x$; $G(\delta R)$ and $R(\delta G)$ are grid-neighborhood support fractions between selected geodesic curves and FTLE-ridge skeletons.}
  \label{tab:geodesic_lcs_timeseries}
  \footnotesize
  \setlength{\tabcolsep}{3.2pt}
  \renewcommand{\arraystretch}{1.18}
  \begin{tabular}{lrrrrrrr}
    \hline\hline
    Family & Samples & Curves & Curves/sample & Median $L_g/L_x$ & IQR $L_g/L_x$ & $G(\delta R)$ & $R(\delta G)$ \\
    \hline
    repelling & 55 & 904 & 16.44 $\pm$ 4.10 & 0.064 & 0.046--0.092 & 0.570 & 0.302 \\
    attracting & 55 & 706 & 12.84 $\pm$ 3.46 & 0.092 & 0.063--0.138 & 0.427 & 0.301 \\
    \hline\hline
  \end{tabular}
\end{table*}

%% file: tables/geodesic_lcs_sensitivity.tex
\begin{table*}[t]
  \centering
  \caption{One-factor sensitivity of planar geodesic-LCS extraction over a matched six-start-time-per-family subset. The baseline uses $p_s=98$, $N_g=24$, and $f_s=0.20$, where $p_s$ is the seed percentile, $N_g$ is the maximum retained curves per time and family, and $f_s$ is the minimum normal-maximum support fraction. Reported values are time means; $G(\delta R)$ and $R(\delta G)$ are grid-neighborhood agreement fractions between geodesic curves and FTLE ridges.}
  \label{tab:geodesic_lcs_sensitivity}
  \footnotesize
  \setlength{\tabcolsep}{3.0pt}
  \renewcommand{\arraystretch}{1.12}
  \begin{tabular}{lllrrrrr}
    \hline\hline
    Variant & Parameter & Family & Samples & Curves/time & $\tilde{L}_g/L_x$ & $G(\delta R)$ & $R(\delta G)$ \\
    \hline
    baseline & reference & repelling & 6 & 13.67 & 0.096 & 0.554 & 0.329 \\
    baseline & reference & attracting & 6 & 11.83 & 0.134 & 0.450 & 0.414 \\
    $p_s=97$ & seed percentile & repelling & 6 & 16.00 & 0.096 & 0.518 & 0.339 \\
    $p_s=97$ & seed percentile & attracting & 6 & 13.67 & 0.133 & 0.443 & 0.432 \\
    $p_s=99$ & seed percentile & repelling & 6 & 10.33 & 0.098 & 0.576 & 0.289 \\
    $p_s=99$ & seed percentile & attracting & 6 & 8.83 & 0.152 & 0.469 & 0.380 \\
    $N_g=16$ & curve cap & repelling & 6 & 12.67 & 0.097 & 0.552 & 0.314 \\
    $N_g=16$ & curve cap & attracting & 6 & 11.67 & 0.134 & 0.454 & 0.414 \\
    $N_g=32$ & curve cap & repelling & 6 & 13.67 & 0.096 & 0.554 & 0.329 \\
    $N_g=32$ & curve cap & attracting & 6 & 11.83 & 0.134 & 0.450 & 0.414 \\
    $f_s=0.15$ & support threshold & repelling & 6 & 13.67 & 0.096 & 0.554 & 0.329 \\
    $f_s=0.15$ & support threshold & attracting & 6 & 11.83 & 0.134 & 0.450 & 0.414 \\
    $f_s=0.25$ & support threshold & repelling & 6 & 13.67 & 0.096 & 0.554 & 0.329 \\
    $f_s=0.25$ & support threshold & attracting & 6 & 11.83 & 0.134 & 0.450 & 0.414 \\
    \hline\hline
  \end{tabular}
\end{table*}

%% file: tables/scalar_coupling_full_timeseries.tex
\begin{table}[t]
  \centering
  \caption{Full-time ridge-conditioned scalar-gradient and projection summaries across all synchronized ridge times.}
  \label{tab:scalar_coupling_full_time}
  \begin{tabular}{lllll}
    \hline
    Direction & Scalar & $E_{\nabla\phi}$ & $P_{|g_{\phi,n}|}$ & Samples \\
    \hline
    forward & $\mathcal{T}$ & 6.338 & 6.216 & 1316703 \\
    forward & $Z$ & 9.534 & 9.614 & 1316703 \\
    forward & $\mathrm{HO_2}$ & 4.087 & 4.105 & 1316703 \\
    backward & $\mathcal{T}$ & 6.536 & 6.432 & 804710 \\
    backward & $Z$ & 9.476 & 9.326 & 804710 \\
    backward & $\mathrm{HO_2}$ & 5.849 & 5.756 & 804710 \\
    \hline
  \end{tabular}
\end{table}

%% file: tables/ridge_scalar_null_significance.tex
\begin{table}[t]
  \centering
  \caption{Time- and cross-stream-stratified null separation for ridge-conditioned scalar-gradient enrichment.}
  \label{tab:ridge_scalar_null}
  \begin{tabular}{lllll}
    \hline
    Direction & Scalar & $R_{obs}$ & $R_{strat}$ (95\% CI) & $R_{obs}/R_{strat}$ \\
    \hline
    forward & $\mathcal{T}$ & 6.338 & 6.020 [6.012,6.028] & 1.053 \\
    forward & $Z$ & 9.534 & 9.086 [9.074,9.099] & 1.049 \\
    forward & $\mathrm{HO_2}$ & 4.087 & 4.085 [4.077,4.092] & 1.000 \\
    backward & $\mathcal{T}$ & 6.536 & 5.662 [5.651,5.673] & 1.155 \\
    backward & $Z$ & 9.476 & 8.046 [8.031,8.061] & 1.178 \\
    backward & $\mathrm{HO_2}$ & 5.849 & 5.193 [5.181,5.205] & 1.126 \\
    \hline
  \end{tabular}
\end{table}

%% file: tables/lead_lag_lcs_metrics.tex
\begin{table}[t]
  \centering
  \caption{Lead/lag summary for detrended FTLE-ridge metric cross-correlations. Positive lag means the first metric leads the second.}
  \label{tab:lcs_lead_lag}
  \begin{tabular}{lllll}
    \hline
    Pair & Peak corr. & Lag [steps] & Lag [iterations] & $N$ \\
    \hline
    $A_f$ vs $I_{fb}$ & 0.933 & 0 & 0 & 541 \\
    $A_b$ vs $I_{fb}$ & 0.838 & 0 & 0 & 541 \\
    $S_f$ vs $I_{fb}$ & -0.489 & -24 & -1200 & 541 \\
    $S_b$ vs $I_{fb}$ & -0.634 & -50 & -2500 & 541 \\
    \hline
  \end{tabular}
\end{table}

%% file: tables/lead_lag_scalar_response.tex
\begin{table}[t]
  \centering
  \caption{Detrended lead/lag summary between intersection activity $I_{fb}$ and scalar enrichment ratios. Positive lag means $I_{fb}$ leads scalar response.}
  \label{tab:scalar_lead_lag}
  \begin{tabular}{llllll}
    \hline
    Direction & Scalar & Peak corr. & Lag [steps] & Lag [iterations] & $N$ \\
    \hline
    forward & $\mathcal{T}$ & -0.662 & 0 & 0 & 541 \\
    forward & $Z$ & -0.693 & 0 & 0 & 541 \\
    forward & $\mathrm{HO_2}$ & -0.703 & 0 & 0 & 541 \\
    backward & $\mathcal{T}$ & -0.685 & 0 & 0 & 541 \\
    backward & $Z$ & -0.709 & 0 & 0 & 541 \\
    backward & $\mathrm{HO_2}$ & -0.694 & 1 & 50 & 541 \\
    \hline
  \end{tabular}
\end{table}

%% file: tables/lcs_geometry_diagnostics.tex
\begin{table}[t]
  \centering
  \caption{FTLE-ridge geometry from forward/backward ridge intersections.}
  \label{tab:lcs_geometry_diag}
  \begin{tabular}{ll}
    \hline
    Metric & Value \\
    \hline
    Mean median crossing angle $\langle\widetilde{\theta}_{fb}\rangle$ & 46.13 \\
    Mean IQR width $\langle\theta_{75}-\theta_{25}\rangle$ & 44.22 \\
    Mean stretching amplification $\left\langle\langle\bar{\sigma}\rangle_{F\cap B}/\langle\bar{\sigma}\rangle_{\Omega}\right\rangle$ & 9.698 \\
    Representative snapshot & mid-sequence \\
    Snapshot median angle $\widetilde{\theta}_{fb}$ & 43.69 \\
    Snapshot stretching amplification & 8.940 \\
    \hline
  \end{tabular}
\end{table}

%% file: tables/lcs_topology_diagnostics.tex
\begin{table}[t]
  \centering
  \caption{FTLE-ridge topology based on forward/backward ridge partitioning over common snapshots.}
  \label{tab:lcs_topology_diag}
  \begin{tabular}{ll}
    \hline
    Metric & Value \\
    \hline
    $\langle A_{F\setminus B}\rangle$ & 8.4962e-03 \\
    $\langle A_{B\setminus F}\rangle$ & 4.8861e-03 \\
    $\langle A_{F\cap B}\rangle$ & 7.8811e-04 \\
    $\langle J\rangle$ & 0.0476 \\
    $\max(J)$ & 0.0932 \\
    $\langle B_{FB}\rangle$ & 0.2003 \\
    $\mathrm{std}(B_{FB})$ & 0.2618 \\
    \hline
  \end{tabular}
\end{table}

%% file: tables/conclusion_metric_sensitivity.tex
\begin{table*}[t]
  \centering
  \caption{Sensitivity of directional transport metrics to available threshold and integration-window variants. Relative change is measured against the reference variant in each group.}
  \label{tab:conclusion_metric_sensitivity}
  \begin{tabular}{lllll}
    \hline
    Group & Metric & Variant & Value & $\Delta$ [\%] \\
    \hline
    Threshold & $R_L=\langle L_f\rangle/\langle L_b\rangle$ & p95 (ref) & 1.894 & 0.00 \\
    Threshold & $R_L=\langle L_f\rangle/\langle L_b\rangle$ & p93 & 2.005 & 5.87 \\
    Threshold & $R_S=\langle S_b\rangle/\langle S_f\rangle$ & p95 (ref) & 1.039 & 0.00 \\
    Threshold & $R_S=\langle S_b\rangle/\langle S_f\rangle$ & p93 & 1.011 & -2.63 \\
    Threshold & $R_{\rho}=\rho_{f,\max}/\rho_{b,\max}$ & p95 (ref) & 1.503 & 0.00 \\
    Threshold & $R_{\rho}=\rho_{f,\max}/\rho_{b,\max}$ & p93 & 1.479 & -1.61 \\
    Window & $R_{\mu}=\langle\sigma_b\rangle/\langle\sigma_f\rangle$ & T=1500 (ref) & 0.7207 & 0.00 \\
    Window & $R_{\mu}=\langle\sigma_b\rangle/\langle\sigma_f\rangle$ & T=1000 & 0.7485 & 3.86 \\
    Window & $R_{p95}=p95(\sigma_b)/p95(\sigma_f)$ & T=1500 (ref) & 0.9565 & 0.00 \\
    Window & $R_{p95}=p95(\sigma_b)/p95(\sigma_f)$ & T=1000 & 0.947 & -1.00 \\
    Window & $R_{\max}=\max(\sigma_b)/\max(\sigma_f)$ & T=1500 (ref) & 1.089 & 0.00 \\
    Window & $R_{\max}=\max(\sigma_b)/\max(\sigma_f)$ & T=1000 & 1.097 & 0.71 \\
    \hline
  \end{tabular}
\end{table*}

%% file: tables/timescale_consistency.tex
\begin{table}[t]
  \centering
  \caption{Timescale-consistency summary for lag interpretation.}
  \label{tab:timescale_consistency}
  \begin{tabular}{ll}
    \hline
    Metric & Value \\
    \hline
    Snapshot spacing $\Delta t$ [iterations] & 50.0 \\
    Integration window $T$ [iterations] & 1500 \\
    Integration window $T$ [samples] & 30.0 \\
    $I_{fb}$ decorrelation scale $\tau_e$ [samples] & 88.0 \\
    $I_{fb}$ decorrelation scale $\tau_e$ [iterations] & 4400.0 \\
    Median $|\ell|$ (ridge metrics) [samples] & 12.0 \\
    Median $|\ell|$ (scalar response) [samples] & 0.0 \\
    Max $|\ell|$ (all reported) [samples] & 50.0 \\
    Fraction with $|\ell|\le \tau_e$ & 1.000 \\
    Fraction with $|\ell|\le T$ & 0.900 \\
    \hline
  \end{tabular}
\end{table}

%% file: tables/uncertainty_envelope_summary.tex
\begin{table*}[t]
  \centering
  \caption{Uncertainty-envelope summary from available threshold and integration-window variants. Relative spread is $(\max-\min)/\mathrm{mid}$ in percent.}
  \label{tab:uncertainty_envelope}
  \footnotesize
  \setlength{\tabcolsep}{3.5pt}
  \begin{tabular}{lllll}
    \hline
    Group & Direction & Metric & Min--Max & Spread [\%] \\
    \hline
    threshold & forward & $\langle L/L_x\rangle$ & $1.78\times10^{-1}$--$2.26\times10^{-1}$ & 23.9 \\
    threshold & backward & $\langle L/L_x\rangle$ & $9.39\times10^{-2}$--$1.13\times10^{-1}$ & 18.2 \\
    threshold & forward & $\langle\sigma_{\mathrm{ridge}}\rangle$ & $7.46\times10^{4}$--$8.16\times10^{4}$ & 8.9 \\
    threshold & backward & $\langle\sigma_{\mathrm{ridge}}\rangle$ & $7.55\times10^{4}$--$8.47\times10^{4}$ & 11.6 \\
    threshold & forward & $\max\,\rho_{\mathrm{ridge}}(x)$ & $5.57\times10^{0}$--$8.48\times10^{0}$ & 41.5 \\
    threshold & backward & $\max\,\rho_{\mathrm{ridge}}(x)$ & $3.70\times10^{0}$--$5.74\times10^{0}$ & 43.1 \\
    window & forward & $\langle\sigma\rangle$ & $1.04\times10^{4}$--$1.20\times10^{4}$ & 13.9 \\
    window & backward & $\langle\sigma\rangle$ & $7.50\times10^{3}$--$8.95\times10^{3}$ & 17.7 \\
    window & forward & $p_{95}(\sigma)$ & $5.91\times10^{4}$--$6.96\times10^{4}$ & 16.4 \\
    window & backward & $p_{95}(\sigma)$ & $5.65\times10^{4}$--$6.59\times10^{4}$ & 15.4 \\
    window & forward & $\sigma_{\max}$ & $1.11\times10^{5}$--$1.46\times10^{5}$ & 26.8 \\
    window & backward & $\sigma_{\max}$ & $1.21\times10^{5}$--$1.60\times10^{5}$ & 27.4 \\
    window & forward & $\mathrm{std}(\sigma)$ & $1.91\times10^{4}$--$2.28\times10^{4}$ & 17.6 \\
    window & backward & $\mathrm{std}(\sigma)$ & $2.12\times10^{4}$--$2.52\times10^{4}$ & 17.3 \\
    \hline
  \end{tabular}
\end{table*}

%% file: tables/ridge_projection_stratified_null.tex
\begin{center}
  \begin{minipage}{\linewidth}
  \captionsetup{type=table}
  \centering
  \captionof{table}{Time- and cross-stream-stratified null separation for ridge-conditioned absolute scalar-gradient projection ratios.}
  \label{tab:projection_stratified_null}
  \scriptsize
  \setlength{\tabcolsep}{1.6pt}
  \begin{tabular}{lllll}
    \hline
    Direction & Scalar & $R_{obs}$ & $R_{strat}$ (95\% CI) & $R_{obs}/R_{strat}$ \\
    \hline
    forward & $\mathcal{T}$ & 6.216 & 5.887 [5.877,5.897] & 1.056 \\
    forward & $Z$ & 9.614 & 9.143 [9.127,9.158] & 1.052 \\
    forward & $\mathrm{HO_2}$ & 4.105 & 4.078 [4.069,4.087] & 1.007 \\
    backward & $\mathcal{T}$ & 6.432 & 5.588 [5.575,5.602] & 1.151 \\
    backward & $Z$ & 9.326 & 7.933 [7.914,7.952] & 1.176 \\
    backward & $\mathrm{HO_2}$ & 5.756 & 5.096 [5.082,5.111] & 1.129 \\
    \hline
  \end{tabular}
  \end{minipage}
\end{center}